\newcommand{\ub}{$\mu_{\text B}$}

\newcommand{\C}{$^\circ$C}
\documentclass{nature}

\usepackage[version=3]{mhchem} % Formula subscripts using \ce{}f
\usepackage[T1]{fontenc}       % Use modern font encodings
\usepackage[version=3]{mhchem} % Formula subscripts using \ce{}
\usepackage[T1]{fontenc}       % Use modern font encodings
\usepackage{graphicx}          % Include figure files
\usepackage{dcolumn}           % Align table columns on decimal point
\usepackage{bm}                % bold math
\usepackage{xcolor}            % text color
\usepackage{colortbl}
\usepackage{amsmath}           % math
\usepackage{hyperref}          % add hypertext capabilities
\usepackage{tabularx}
\usepackage{lineno}
\usepackage{chemformula} % Formula subscripts using \ch{}
\usepackage{xcolor,colortbl}
\usepackage{multirow}
\usepackage{booktabs}
\title{
\begin{center}
{Magnetoresistive Memory in the Paramagnetic Phase of \ch{Eu5In2As6}}
\end{center}
}
\author{
Sudhaman~R.~Balguri$^{1}$,
Mira~B.~Mahendru$^{1}$,
Rourav~Basak$^{2}$,
Enrique~O.~González-Delgado$^{1}$,
Adam~A.~Aczel$^{3}$,
David~E.~Graf$^{4}$,
Andreas~Rydh$^{5}$,
Christopher~C.~Homes$^6$,
Jonathan~Gaudet$^{7,8}$,
Ying~Ran$^1$,
Alex~Frano$^{2,9}$, and
Fazel~Tafti$^1$
}
\begin{document}
\maketitle

\begin{affiliations}
 \item{Department of Physics, Boston College, Chestnut Hill, MA 02467, USA}
 \item{Department of Physics, University of California San Diego, California 92093, USA}
 \item{Neutron Scattering Division, Oak Ridge National Laboratory, Oak Ridge, TN 37831, USA}
 \item{National High Magnetic Field Laboratory, Tallahassee, Florida 32310, USA}
 \item{Department of Physics, Stockholm University, SE-10691 Stockholm, Sweden}
 \item{National Synchrotron Light Source II, Brookhaven National Laboratory, Upton, New York 11973, USA}
 \item{NIST Center for Neutron Research, National Institute of Standards and Technology, Gaithersburg, Maryland 20899, USA}
 \item{Department of Materials Science and Engineering, University of Maryland, College Park, MD 20742-2115, USA}
 \item{Program in Materials Science and Engineering, University of California San Diego, California 92093, USA} 
\end{affiliations}

%%%%%%%%%%%%%%%%%%%%%%%%%%%%%%%%%%
%%%%%%%%%%%%% ABSTRACT %%%%%%%%%%%
%%%%%%%%%%%%%%%%%%%%%%%%%%%%%%%%%%
% \pagebreak
\begin{abstract}
Magnetoresistive materials that respond sensitively to applied fields are central to modern data storage technologies.
Here we unveil a novel Magnetoresistive Memory (MRM) in \ch{Eu5In2As6}, where the electrical resistivity depends not only on the magnitude but also on the history of the applied magnetic field.
Such an effect has been reported in only two classes of strongly correlated electron systems: perovskite manganites~\cite{dagotto_colossal_2001,uehara_percolative_1999,gordon_temperature_2001,kuwahara_first-order_1995,tomioka_collapse_1995,yoshizawa_neutron-diffraction_1995,lai_mesoscopic_2010,levy_novel_2002} and pyrochlore iridates~\cite{ueda_anomalous_2014,tian_field-induced_2016,matsuhira_metalinsulator_2011,arima_time-reversal_2013,ma_mobile_2015}.
In both cases, the effect has been observed in the magnetically ordered phase.
It has been attributed to metastable magnetic states in manganites and conducting domain walls in iridates.
Remarkably, the MRM in \ch{Eu5In2As6} onsets at twice the antiferromagnetic transition temperature, well within the paramagnetic phase.
The temperature, field, and time dependence of resistivity suggest that either a hidden order or a fluctuating phase with short-range correlations underlies this effect. 
Our results offer MRM as a new platform for quantum sensing and memory technologies, and encourage searching for MRM in related materials.
\end{abstract}

%%%%%%%%%%%%%%%%%%%%%%%%%%%%%%%%%%
%%%%%%%%%%%%%% BODY %%%%%%%%%%%%%%
%%%%%%%%%%%%%%%%%%%%%%%%%%%%%%%%%%
% \pagebreak
% \linenumbers
%%%%%%%%%%\linenumbers
% Then the body of the main text appears after the intro paragraph.
% Figure environments can be left in place in the document.
% \verb|\includegraphics| commands are ignored since Nature wants
% the figures sent as separate files and the captions are
% automatically moved to the end of the document (they are printed
% out with the \verb|\end{document}| command. However, tables must
% be manually moved to the end of the document, after the addendum.

\section*{\label{sec:intro}Introduction}
Magnetoresistance (MR) is a powerful probe of electronic structure~\cite{collaudin_angle_2015,grissonnanche_linear-temperature_2021,yang_extreme_2017, ali_large_2014,suzuki_singular_2019} and the basis of magnetic sensing and memory devices
~\cite{reig_magnetic_2009,kao_unconventional_2025}.
Colossal Magnetoresistance (CMR) is a negative MR characterized by orders-of-magnitude drop in the resistivity under an applied magnetic field, which results from the interplay of spin, charge, and lattice degrees of freedom.~\cite{dagotto_colossal_2001} 
In rare cases, materials with a CMR also exhibit a dependence of their resistivity on the history of the applied magnetic field. 
This effect, dubbed a magnetoresistive memory (MRM) here, enables the advancement of next-generation magnetic field-effect memristors for neuromorphic computing.~\cite{marti_room-temperature_2014,liu_cryogenic_2025}

So far, MRM has been reported in two classes of CMR materials: the perovskite manganites and pyrochlore iridates.
Manganites have a mixed Mn$^{3+}$/Mn$^{4+}$ valence, sizable Jahn-Teller distortion, and large Hund's coupling (2-3 eV), leading to competing interactions~\cite{dagotto_colossal_2001}.
Upon zero-field-cooling (ZFC), such materials can settle in a metastable state comprising both ferromagnetic (FM) metallic and charge-ordered antiferromagnetic (AFM) insulating phases~\cite{dagotto_colossal_2001,uehara_percolative_1999}.
An applied magnetic field can train the metastable state into a stable state by collapsing the charge order and favoring the metallic FM order~\cite{gordon_temperature_2001,kuwahara_first-order_1995,tomioka_collapse_1995,yoshizawa_neutron-diffraction_1995,lai_mesoscopic_2010,levy_novel_2002}.

%%%%%%%%%%%%%%%%%% FIGURE1 %%%%%%%%%%%%%%%%%%%
\begin{figure*}
  \includegraphics[width=\textwidth]{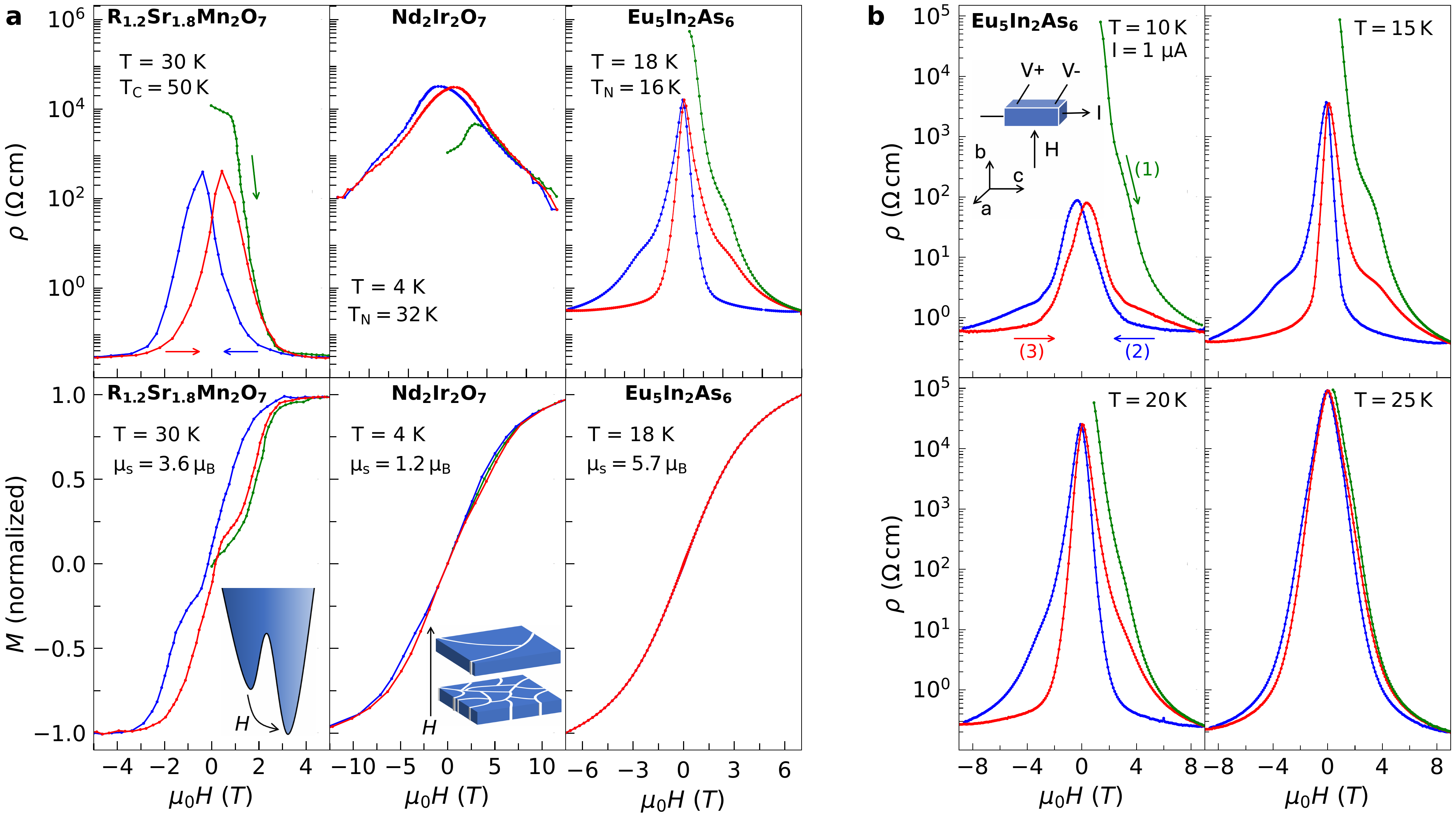}
      \caption{\label{fig:RH_merged_S12}
 \textbf{Field dependence of MRM.}
 (a) Comparing MRM between (left) the manganite {R}$_{1.2}${Sr}$_{1.8}${Mn}$_2${O}$_7$ with $R = \mathrm{La_{0.4}Pr_{0.6}}$~\cite{gordon_temperature_2001}, (middle) the iridate  \ch{Nd2Ir2O7}~\cite{ueda_anomalous_2014}, and (right) \ch{Eu5In2As6}.
 The top and bottom panels show the resistivity and magnetization data, respectively. 
 The green, blue, and red curves represent ZFC, downsweep and upsweep data. 
 Insets illustrate field training from a metastable to a stable state in the manganite, and field training of conducting domain walls in the iridate. 
 All $M(H)$ curves are normalized to the saturated moment $\mu_s$.
 (b)  Field dependence of resistivity in \ch{Eu5In2As6} at several temperatures below and above $T_\mathrm{N}=16$~K. 
 The field sweep rate is 100~Oe/s, and similar results are obtained at half or twice this rate.
 The numbers indicate the order of ZFC (green), downsweep (blue), and upsweep (red) measurements. 
 The resistivity exceeded the lock-in detection limit and was not measurable below $H = 1.5~\mathrm{T}$ in the untrained state (ZFC).
 }
\end{figure*}
%%%%%%%%%%%%%%%%%%%%%%%%%%%%%%%%%%%%%%%%%%%%%%%

An example of MRM in manganites is shown in Fig.~\ref{fig:RH_merged_S12}a for ({La}$_{0.4}${Pr}$_{0.6}$)$_{1.2}${Sr}$_{1.8}${Mn}$_2${O}$_7$ below its field-induced FM transition at $T_{\text C}=50$~K~\cite{gordon_temperature_2001}.
The ZFC resistivity and magnetization data (green curves in Fig.~\ref{fig:RH_merged_S12}) are obtained by first cooling the sample in zero field, then measuring $\rho(H)$ and $M(H)$ as the field is increased from 0 to 5~T.
When the field is subsequently reduced from 5~T back to zero (blue curves), both $\rho(H)$ and $M(H)$ follow new traces without overlapping the ZFC data.
Notably, $\rho(H=0)$ decreases by two orders of magnitude after field training, due to growing FM metallic domains.
Both the resistivity and magnetization curves follow different traces during $+5\to-5$~T (blue) and $-5\to+5$~T (red) field sweeps. 
Thus, the material keeps a memory of its exposure to the magnetic field.

MRM has also been reported in the iridate system \ch{Nd2Ir2O7} at temperatures below its AFM transition, $T_\mathrm{N} = 32\,\mathrm{K}$~\cite{ueda_anomalous_2014,tian_field-induced_2016}.
As shown in Fig.~\ref{fig:RH_merged_S12}a (middle panel), the ZFC resistivity and magnetization curves (green) differ from the subsequent field sweeps (blue and red). 
For consistency, we display the downsweep data in blue and the upsweep data in red across all materials.
The non-overlapping downsweep and upsweep curves indicate a broken time-reversal symmetry (TRS), as expected below the magnetic transition temperature in both manganites and iridates. 

Unlike in manganites, where $\rho(H=0)$ decreases upon field training, \ch{Nd2Ir2O7} exhibits an order of magnitude increase in $\rho(H=0)$, implying a different mechanism for MRM.
\ch{Nd2Ir2O7} transitions from a paramagnetic metal to an AFM insulator below $T_\mathrm{N}$~\cite{matsuhira_metalinsulator_2011,arima_time-reversal_2013}.
Two domains of all-in–all-out (AIAO) order emerge at $T<T_\text{N}$, and the domain walls separating them hold a paramagnetic metallic state.
Field training yields larger magnetic domains and fewer conducting domain walls (inset of Fig.~\ref{fig:RH_merged_S12}a), hence higher resistivity~\cite{ma_mobile_2015,ueda_anomalous_2014}. 
The AIAO AFM order breaks TRS, explaining the asymmetry between the upsweep and downsweep curves in Fig.~\ref{fig:RH_merged_S12}a~\cite{tian_field-induced_2016,arima_time-reversal_2013}.

The above MRM materials, manganites and iridates, share two key features.
Both are strongly correlated systems with metal-insulator transitions, and both exhibit MRM only when a long-range magnetic order breaks TRS~\cite{tomioka_collapse_1995,matsuhira_metalinsulator_2011,ueda_anomalous_2014}.
In this article, we report MRM in a new regime by focusing on \ch{Eu5In2As6}, a weakly correlated semiconductor devoid of metal-insulator transitions, that exhibits MRM even in its paramagnetic phase.

\section*{\label{sec:results}Results}
The right column in Fig.~\ref{fig:RH_merged_S12}a shows a pronounced MRM in \ch{Eu5In2As6} similar to those of manganites and iridates, but with two key differences. 
First, the effect is observed above the AFM transition temperature ($T_\mathrm{N}=16$~K).
Second, it has no counterpart in the magnetization data. 
These observations show that MRM extends well into the paramagnetic (PM) phase of \ch{Eu5In2As6}, unlike in manganites and iridates where the effect is restricted to the magnetically ordered phase.
Therefore, metastable magnetic states (as in manganites) or magnetic domain walls (as in iridates) cannot explain the MRM effect in the title compound.
The non-overlapping upsweep and downsweep curves indicate a broken TRS in the absence of a long-range magnetic order.
This could result from either a short-range correlated state, such as magnetic polarons that extend into the PM phase~\cite{balguri_two_2025}, or a long-range hidden order ~\cite{chandra_hidden_2002,haule_arrested_2009,patri_unveiling_2019,onimaru_antiferroquadrupolar_2011,tokunaga_nmr_2006,aynajian_visualizing_2010} at $T>T_\mathrm{N}$.

% These observations discredit a purely magnetic mechanism for the MRM in \ch{Eu5In2As6}, unlike in manganites (metastable magnetic states) and iridates (conducting domain walls). 
% The non-overlapping upsweep and downsweep curves indicate a broken TRS in the absence of a long-range magnetic order at $T>T_\text{N}$.
% This is possible if some hidden order ~\cite{chandra_hidden_2002,haule_arrested_2009,patri_unveiling_2019,onimaru_antiferroquadrupolar_2011,tokunaga_nmr_2006,aynajian_visualizing_2010} breaks the TRS above \TN.

\ch{Eu5In2As6} is a CMR-exhibiting magnetic semiconductor with a band gap of 45~meV (supplementary Fig.~S1) and an A-type AFM transition at $T_\mathrm{N}=16$~K~\cite{balguri_two_2025,varnava_engineering_2022}.
It has a carrier concentration of $10^{18}$~cm$^{-3}$~(Fig.~S1), orders of magnitude smaller than $10^{21}$~cm$^{-3}$ in the metallic states of manganites and iridates~\cite{dagotto_colossal_2001,disseler_magnetization_2013}.
Thus, the transport properties of this compound, including CMR and MRM, are attributed to a small concentration of extrinsic carriers rather than a coherent Fermi surface.
In the following sections, we will examine the field, temperature, and time dependence of MRM. %, which collectively suggest that a hidden order may underlie the MRM in this material.

\section*{\label{sec:field}Field Dependence of MRM}
%%%%%%%%%%%%%%%%%% FIGURE2 %%%%%%%%%%%%%%%%%%%
\begin{figure*}
  \includegraphics[width=\textwidth]{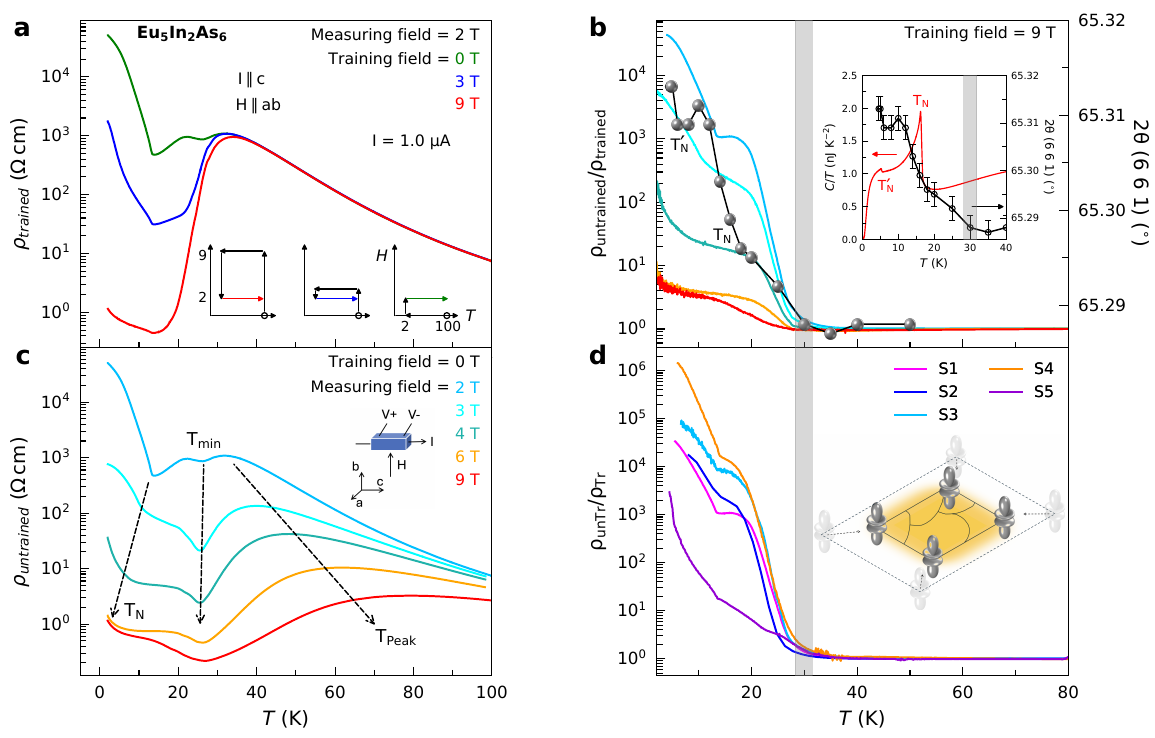}
      \caption{\label{fig:RT_merged_S12}
 \textbf{Temperature dependence of MRM.}
 (a) Three resistivity curves measured from 2 to 100 K under the same measuring field of 2~T, after the sample was cooled under a training field of 0~T (green), 3~T (blue), and 9~T (red).
 The inset shows the training procedure (black arrows) before measuring (colored arrows); the circles mark the starting point.
 (b) MRM appears below 30~K, where the ratio of untrained to trained resistivity diverges.
 The gray spheres mark the position of the (661) Bragg peak.
 The inset shows that no anomalies are observed in the specific heat at 30~K. 
 (c) Three features in $\rho_\text{untrained}(T)$ include $T_{\mathrm{peak}}$ (broad maximum due to polaron percolation), $T_\mathrm{N}$ (resistivity upturn due to AFM ordering),  and $T_{\mathrm{min}}$ (resistivity minimum due to hybridization). %coinciding with the onset of MRM).
 (d) MRM starts at the same temperature in different samples. The inset illustrates the simultaneous occurrence of MRM and lattice contraction, both of which could be due to hybridization between $f$ electrons (gray) and extrinsic carriers (yellow). %, leading to a hidden order with domain walls.   
 Measurements were performed on sample S1 with current along the $c$-axis and field in the $ab$-plane, same as in Fig.~\ref{fig:RH_merged_S12}.}
 
\end{figure*}
%%%%%%%%%%%%%%%%%%%%%%%%%%%%%%%%%%%%%%%%%%%%%%% 
Figure~\ref{fig:RH_merged_S12}b shows the field dependence of resistivity at a few temperatures below and above $T_\mathrm{N}$.
The $\rho(H)$ curves at 10~K exhibit two characteristic features of MRM.
First, $\rho(H=0)$ is three orders of magnitude larger in the ZFC curve (green) than in subsequent field sweeps.
Second, the downsweep (blue) and upsweep (red) curves do not overlap.
The first feature is irreversible and disappears after the first field sweep. 
In contrast, the second feature is reversible, i.e. all subsequent downsweep and upsweep curves trace the same hysteresis loop.
These features are reproducible across different samples (Fig.~S2).
% \textcolor{blue}{The resistivity data presented in the main text corresponds to S1, and its magnetization data are shown in Fig.S10. In Fig.S4 we present the angle dependence of this effect, which no significant changes.} 
For consistency, all transport data presented in this article are from sample S1 in the configuration shown in the inset of Fig.~\ref{fig:RH_merged_S12}b.
The MRM effect shows little dependence on the angle between current and field with respect to the sample, as shown in Fig.~S3.

The gap between the ZFC data (green) and the subsequent field sweeps (blue/red) becomes smaller but does not vanish as the temperature increases above $T_\mathrm{N}=16$~K (Fig.~\ref{fig:RH_merged_S12}b).
For example, at 20~K, there is still an order of magnitude difference between the green and red/blue curves near $H=0$. 
The hysteresis loop between the upsweep and downsweep curves also persists up to 25 K, well into the PM phase. 
This is in stark contrast to manganites and iridates that do not exhibit MRM above their magnetic ordering temperatures. 
% Thus, the MRM in \ch{Eu5In2As6} could be related to TRS breaking of a higher order than a static dipolar magnetic order. 

\section*{\label{sec:temperature}Temperature Dependence of MRM}
To explain the temperature dependence of MRM, we need to define the \textit{training field} and \textit{measuring field}.
The former refers to the field under which the sample is cooled to 2~K prior to data collection.
The latter is the field at which the $\rho(T)$ curve is measured from 2 to 100~K.
The inset of Fig.~\ref{fig:RT_merged_S12}a illustrates how the sample is prepared (trained) for the measurement.
All three curves in Fig.~\ref{fig:RT_merged_S12}a are measured from 2 to 100~K under a \emph{measuring field} of 2~T.
But before each measurement, the sample was cooled under a different \emph{training field}: 0~T (green), 3~T (blue), and 9~T (red).
We refer to these curves as \emph{untrained}, \emph{weakly trained}, and \emph{trained}, respectively.

It is no surprise that the three curves in Fig.~\ref{fig:RT_merged_S12}a overlap at $T>30$~K, since they were all measured on the same sample, from 2 to 100~K, under the same measuring field of 2~T.
The striking observation, and the hallmark of MRM, is the divergence of these curves at $T<30$~K.
Specifically, the resistivity value at 2~K is four orders of magnitude larger in the untrained curve (green) than in the trained one (red), while the weakly trained curve (blue) falls in between.
The material keeps a memory of how it was prepared for the measurement, effectively acting as a magnetic field recorder. 

The onset of MRM near 30~K is highlighted in Fig.~\ref{fig:RT_merged_S12}b by plotting the ratio $\rho_\text{untrained}/\rho_\text{trained}$ as a function of temperature under different measuring fields.
For example, the 2~T curve in Fig.~\ref{fig:RT_merged_S12}b is produced by dividing the green data by the red data in Fig.~\ref{fig:RT_merged_S12}a.
The complete set of $\rho_\text{untrained}(T)$ and $\rho_\text{trained}(T)$ data is presented in supplementary Fig.~S4.
All curves in Fig.~\ref{fig:RT_merged_S12}b collapse to unity at $T>30$~K, since $\rho_\text{untrained}=\rho_\text{trained}$ at high temperatures (Fig.~\ref{fig:RT_merged_S12}a).
However, at $T<30$~K, $\rho_\text{untrained}\neq\rho_\text{trained}$, so the ratio diverges by orders of magnitude.
% The order-parameter-like rise of $\rho_\text{untrained}/\rho_\text{trained}$ in Fig.~\ref{fig:RT_merged_S12}b signals the onset of a hidden order at about 30~K, nearly 2\TN.

The order-parameter-like rise of $\rho_\text{untrained}/\rho_\text{trained}$ in Fig.~\ref{fig:RT_merged_S12}b can be interpreted as the onset of a long-range order at about 30~K, nearly $2T_\mathrm{N}$.
However, the specific heat data in the inset of Fig.~\ref{fig:RT_merged_S12}b do not show an anomaly near 30~K, which means if such a hidden order exists, it must have a small entropic footprint buried in the phonon contribution.
This is possible if the hidden order is restricted to electronic degrees of freedom (orbital or multipolar).
Since \ch{Eu5In2As6} has a small carrier concentration of about 10$^{18}$~cm$^{-3}$ (Fig.~S1), it is likely that the phonon contribution to specific heat overwhelms the electronic contribution.
Note that we use the term ``hidden order'' in a different sense than used for materials such as \ch{URu2Si2} that exhibit a peak in the specific heat without anomalies in the magnetization or transport data~\cite{chandra_hidden_2002,haule_arrested_2009,aynajian_visualizing_2010}.
In \ch{Eu5In2As6}, an anomaly is observed in transport data, but not in the specific heat or magnetization.
% This interpretation is reinforced by the fact that all samples that show MRM have the same onset temperature (30~K) as shown in \textcolor{cyan}{Fig.~S5?}.

% To probe this hidden order, we compare $\rho_\text{untrained}$ as a function of temperature to $\rho_\text{untrained}/\rho_\text{trained}$ at a few select fields in Figs.~\ref{fig:RT_merged_S12}b,c.
The untrained resistivity curves in Fig.~\ref{fig:RT_merged_S12}c exhibit three characteristic features:
(i) A broad peak at $T_{\mathrm{peak}}$ due to the formation and percolation of magnetic polarons -- short-range FM islands in the paramagnetic phase of \ch{Eu5In2As6} and related compounds~\cite{balguri_two_2025,rosa_colossal_2020,Crivillero2023,kopp_robust_2026}.
$T_{\mathrm{peak}}$ increases with increasing field, since polarons form more easily in a magnetic field.
(ii) An upturn due to AFM ordering at $T_\mathrm{N}$, which decreases with increasing field.
(iii) A resistivity minimum at $T_{\mathrm{min}}$, which is nearly field-independent.
Features (i) and (ii) correspond to two types of CMR discussed in an earlier work~\cite{balguri_two_2025} (see also Fig.~S4).
% Feature (iii), however, coincides with the onset of the hidden order around 30~K (shaded area in Figs.~\ref{fig:RT_merged_S12}b,c).
Feature (iii), the resistivity minimum at $T_{\mathrm{min}}$, is close to but slightly below the onset temperature of MRM. Importantly, it does not change under different measuring fields, unlike  $T_\mathrm{N}$ and $T_{\mathrm{peak}}$, suggesting a correlation between the resistivity minimum and MRM.

A resistivity minimum is the signature of the Kondo effect in metals.
But \ch{Eu5In2As6} is a semiconductor and lacks the underlying Fermi surface in a typical Kondo system, especially those with a hidden order~\cite{chandra_hidden_2002,haule_arrested_2009,aynajian_visualizing_2010}.
Nevertheless, some form of hybridization between the Eu $f$-states and the extrinsic carriers in \ch{Eu5In2As6} could be responsible for the resistivity minimum at $T_{\mathrm{min}}$.
Consistent with this picture, both MRM and resistivity minimum are absent in the sister compound \ch{Eu5In2Sb6} (Fig.~S5), suggesting that the presence of one relies on the other.

Whether such a hybridization is responsible for a hidden electronic order requires more detailed experimental investigations.
But an important observation that suggests MRM is an intrinsic effect, rather than due to sample inhomogeneity or surface effect, is the similar onset of MRM in different samples.
The five samples in Fig.~\ref{fig:RT_merged_S12}d have varying carrier densities between $10^{17}$~cm$^{-3}$ and $10^{18}$~cm$^{-3}$, but they all exhibit MRM at the same temperature.
% As explained in the Supplementary Information (\textcolor{cyan}{Fig.~S1 and Table~S1}), samples with carrier densities below the threshold of $10^{17}$~cm$^{-3}$ do not exhibit MRM altogether.
% These observations suggest that a minimum carrier density is required for both the hybridization and MRM to occur.

\section*{\label{sec:xrd}Magnetoelastic Coupling and MRM}
%%%%%%%%%%%%%%%%%% FIGURE3 %%%%%%%%%%%%%%%%%%%
\begin{figure*}
  \includegraphics[width=\textwidth]{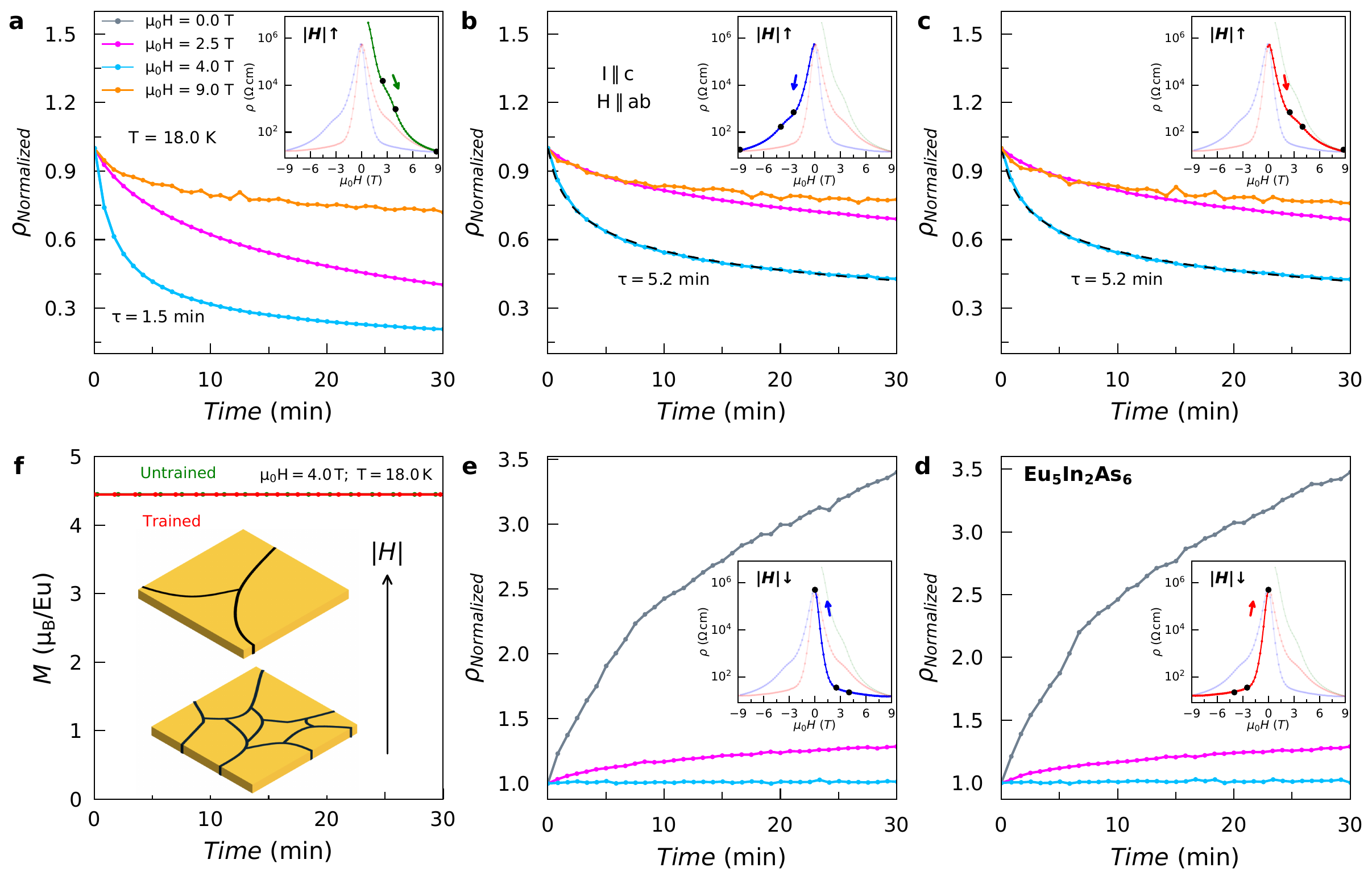}
      \caption{\label{fig:RTime_merged_S12}
 \textbf{Time dependence of MRM.}
 Normalized resistivity at $T=18$~K as a function of time at a few fixed fields, measured during (a) increasing field after the initial ZFC, (b) increasing field when $H<0$, (c) increasing field when $H>0$, (d) decreasing field when $H<0$, (e) decreasing field when $H>0$. Insets show the field sweep directions.
 (f) Time dependence of magnetization at 18~K showing no variation with time. The graphic illustrates the training of hidden-order domain walls (source of scattering) during the field-increasing process. These measurements were performed on the same sample S1 used in Fig.~\ref{fig:RH_merged_S12} and~\ref{fig:RT_merged_S12}.}
\end{figure*}
%%%%%%%%%%%%%%%%%%%%%%%%%%%%%%%%%%%%%%%%%%%%%%%
% There are no anomalies in either the magnetization or specific heat at the onset of hidden order in \ch{Eu5In2As6} (Fig.~\ref{fig:RT_merged_S12}b, inset).
There are no anomalies in either the magnetization or specific heat at 30~K where MRM appears in \ch{Eu5In2As6} (Fig.~\ref{fig:RT_merged_S12}b, inset).
However, x-ray data in Fig.~\ref{fig:RT_merged_S12}b reveal an accelerated lattice contraction below 30~K where MRM begins.
The gray spheres show that the (661) Bragg peak shifts to higher angles, hence lattice contraction, as the temperature is reduced below 30~K, where both MRM and hybridization ($T_{\mathrm{min}}$) start. %$T_{\mathrm{min}}$.
% This implies that the hybridization between $f$ moments and conduction electrons at the onset of hidden order, reflected in the resistivity minimum, involves not only charge and spin, but also lattice degrees of freedom (Fig.~\ref{fig:RT_merged_S12}, graphic).
This implies that both MRM and hybridization between $f$ moments and conduction electrons involve not only charge and spin, but also lattice degrees of freedom, as illustrated in the inset of Fig.~\ref{fig:RT_merged_S12}d.

There are three changes of slope in the temperature dependence of the (661) Bragg peak in Fig.~\ref{fig:RT_merged_S12}b.
The first one is near the onset of MRM at 30~K. %$T_\text{min}=30$~K.
The second one is near the onset of AFM order at $T_\mathrm{N}=16$~K , and the third one is near a spin reorientation transition at $T'_{\mathrm{N}}=6$~K~\cite{balguri_two_2025}.
% These features suggest that a strong magnetoelastic coupling is involved in establishing the hidden order and MRM.
These features suggest that a strong magnetoelastic coupling may be involved in establishing MRM.
A more detailed analysis of the lattice parameters, using both (661) and (660) peaks, is presented in Fig.~S7.

\section*{\label{sec:time}Time Dependence of MRM}
% Like other types of order, the hidden order could exhibit a domain structure.
% Training the sample in a magnetic field would reduce the number of domain walls \textcolor{purple}{associated with the hidden order,} and decrease electron scattering from them, hence reducing resistivity upon field training (Figs.~\ref{fig:RH_merged_S12} and \ref{fig:RT_merged_S12}).
% This is the opposite scenario to the conducting domain walls in \ch{Nd2Ir2O7}, where field training removes them and enhances the resistivity (Fig.~\ref{fig:RH_merged_S12}a, middle panel).
% In this subsection, we provide evidence of domain wall scattering in \ch{Eu5In2As6} from the time dependence of resistivity.

Concurrent with the onset of MRM below 30~K, the electrical resistivity of \ch{Eu5In2As6} exhibits a curious time dependence. 
This is seen in Fig.~\ref{fig:RTime_merged_S12}, where resistivity is traced as a function of time at 18~K ($T_\mathrm{N}<T<T_{\mathrm{min}}$) and at a few select fields.
The insets specify during which part of the field sweep (increasing or decreasing, positive or negative), the data were collected.
Each curve is normalized to its $t = 0$ value to account for the large CMR.

After zero-field cooling to 18~K, we increase the field to 2.5~T and trace the time dependence of resistivity at that field for 30 minutes (Fig.~\ref{fig:RTime_merged_S12}a), then repeat the process at 4~T and 9~T.
We observe a relaxor-like behavior at all three fields, but it is most pronounced at 4~T. 
Fitting the time-dependence at 4~T to a stretched exponential function, $\rho=\rho_0\left[1-\exp(-\tau/t)^\beta\right]$, yields a relaxation time $\tau=1.5$~min. 
This behavior is reproduced in subsequent field sweeps when the field magnitude is increased, regardless of its sign (Figs.~\ref{fig:RTime_merged_S12}b,c), albeit with a longer relaxation time, $\tau=5.2$~min.
The stretched exponential fit is a phenomenological model based on the assumption that charge carriers diffuse through a random potential in an ordered state~\cite{imry_random-field_1975}. 
The random potential could be due to either domain walls (DW) of a hidden order or the presence of magnetic polarons.

Similar time-dependent measurements in the field-decreasing parts of the cycle (Figs.~\ref{fig:RTime_merged_S12}d,e) reveal a resistivity enhancement, rather than relaxation.
Thus, the time dependence of resistivity, at the same field, depends on whether that field was attained during a field-increasing or field-decreasing process. 

This observation can be explained in terms of DW scattering.
Increasing the field magnitude, regardless of its sign, leads to growing domains, hence less DW scattering.
However, this process can be retarded, such that when the field is paused, the domains keep growing, DWs keep disappearing, and resistivity keeps relaxing with a characteristic timescale $\tau$.
Conversely, decreasing the field magnitude, regardless of its sign, leads to a retarded proliferation of domains and DWs, hence a delayed resistivity enhancement.

The shorter relaxation rate in the initial field sweep (ZFC data in Fig.~\ref{fig:RTime_merged_S12}a) is consistent with the distinction we made earlier between irreversible and reversible processes.
The first cycle (ZFC) likely involves an irreversible initial training of DWs, while the consistent behavior in subsequent field sweeps reflects a reversible response.
The resistivity relaxation vanishes above 30~K along with MRM (Fig.~S4), supporting the interpretation that the field training of these DWs underlies the MRM.

Resistivity relaxation has also been reported in a few manganites due to competing phases and spin-glass magnetism.~\cite{matsukawa_resistive_2005,gordon_temperature_2001,levy_novel_2002,nam_ferromagnetism_2000}.
Unlike manganites, \ch{Eu5In2As6} shows neither a time-dependent magnetization (Fig.~\ref{fig:RTime_merged_S12}f) nor a frequency-dependent AC susceptibility (Fig.~S6), ruling out a spin-glass mechanism. However, short-range magnetic polarons could provide a mechanism for resistivity relaxation~\cite{matsukawa_resistive_2005}. 
The relaxor-like behavior is independent of the applied current, ruling out Joule heating effects (Fig.~S6).

\section*{\label{sec:discussion}Discussion}
Our experimental results reveal a novel MRM effect that starts in the paramagnetic phase of \ch{Eu5In2As6}, unlike the memory effects in manganites and iridates that appear only below the magnetic transition temperature.
MRM is marked by orders of magnitude increase in $\rho_\text{untrained}/\rho_\text{trained}$ below 30~K, nearly twice the $T_\mathrm{N}=16$~K .
At this temperature, we do not find an anomaly in the specific heat, but we find an accelerated lattice contraction rate, and a time dependence of resistivity below 30~K.
These observations can be explained either by domain wall dynamics of a hidden order or by a short-range correlated phase such as magnetic polarons. 
While we cannot distinguish these two mechanisms with certainty, we discuss the evidence for and against each mechanism.

Magnetic polarons emerge above $T_\mathrm{N}$, in the paramagnetic phase of \ch{Eu5In2As6} and related materials such as \ch{EuCd2P2}~\cite{balguri_two_2025,kopp_robust_2026,li_colossal_2024} and \ch{Eu5In2Sb6}~\cite{rosa_colossal_2020,Crivillero2023,morano_noncollinear_2024}. 
They are FM islands within the PM phase; these polarons could undergo different percolation patterns under different training fields, resulting in different resistive states leading the the observed hysteresis loops. Magnetic polarons could also explain the time-dependent of resistivity, according to the models proposed for manganites~\cite{matsukawa_resistive_2005}. 
Being short-range correlated islands, they do not contradict the absence of thermodynamic anomalies at 30~K, where MRM onsets.

Although a short-range correlated state offers a plausible explanation for MRM, at least two observations challenge this scenario.
First, $T_{\mathrm{peak}}$, the characteristic temperature for polaronic percolation, increases with increasing magnetic field (Fig.~\ref{fig:RT_merged_S12}c), unlike the field-independent onset of MRM.
For example, $T_\text{peak}=55$~K at 6~T (Fig.~\ref{fig:RT_merged_S12}c), whereas MRM starts below 30~K. 
Second, the MRM persists even at 9~T, with the $\rho_\text{untrained}/\rho_\text{trained}$ ratio reaching a value of 10 at 2~K, despite the fact that, at 2~K, a full spin polarization of 7~\ub\ is reached by 6~T~\cite{balguri_two_2025}.

We now turn to hidden order as an alternative explanation for the MRM and time dependence of resistivity below 30~K due to dynamical domain wall scattering.
The resistivity minimum at $T_{\mathrm{min}}$ suggests that an electronic reconstruction due to $s$-$f$ or $p$-$f$ hybridization is responsible for such an ordered state.
A long-range hidden order is also consistent with the same onset temperature for the MRM across samples with different carrier concentrations.

The main observation against a hidden order is the lack of a specific heat anomaly at 30~K (inset of Fig.~\ref{fig:RT_merged_S12}b).
One explanation is that phonons can mask the specific heat signature of a fragile electronic order in a system with low carrier concentrations.
AC elastocaloric effect could be a good experimental probe for such electronic transitions, especially because it admits little contribution from phonons~\cite{ikeda_ac_2019}.
The lattice contraction at the onset of MRM suggests that magnetostriction can also be a good probe of the hidden order.
The angle dependence of magnetostriction is sensitive to the underlying symmetries of a hidden order~\cite{patri_unveiling_2019}.  

%%%%%%%%%%%%%%%%%% FIGURE4 %%%%%%%%%%%%%%%%%%%
\begin{figure*}
 \centering
  \includegraphics[width=0.6\textwidth]{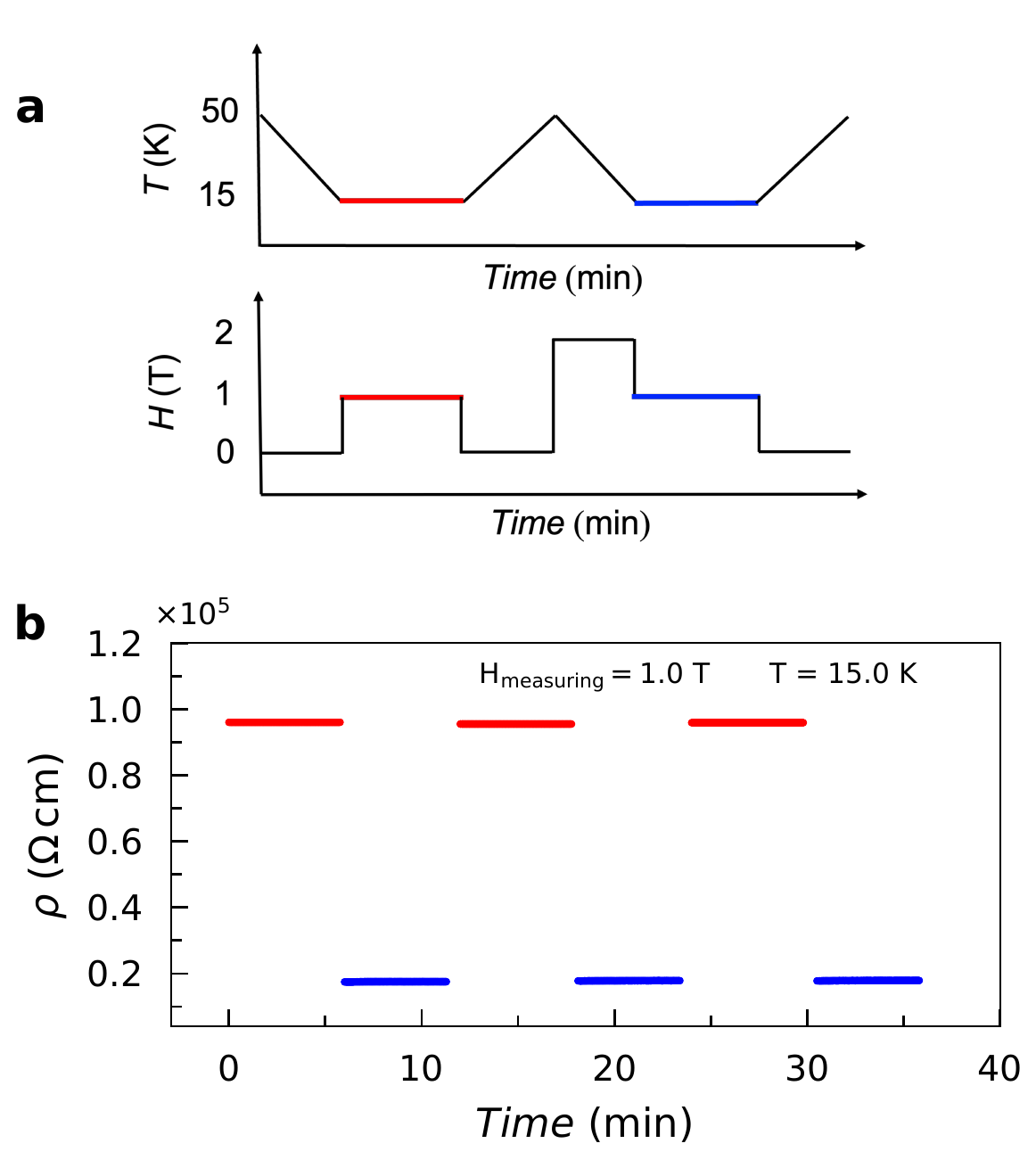}
      \caption{\label{fig:Applicaton}
 \textbf{Application of MRM.}
 (a) A training sequence, where black lines show the training process and the colored lines show the measurement process. (b) The red data were obtained from an untrained \ch{Eu5In2As6} crystal measured at 1~T, while the blue data were obtained from the same crystal after training under 2~T and subsequent measuring at 1~T.
The training protocol is repeated three times, showing reproducible resistive states.}
\end{figure*}
%%%%%%%%%%%%%%%%%%%%%%%%%%%%%%%%%%%%%%%%%%%%%%%

Regardless of its origin, MRM is an applicable effect that enables engineering well-defined resistive states through specific training and measuring protocols.
For example, Fig.~\ref{fig:Applicaton} illustrates two distinct resistive states in \ch{Eu5In2As6}, marked by red and blue colors. 
The red curve corresponds to the untrained states measured at 1~T, whereas the blue curve represents the trained state, obtained by training under 2~T and measuring at 1~T. 
Although both states are measured at the same temperature (15~K) and field (1~T), they exhibit distinct resistivity values arising solely from their different magnetic field histories. 
We repeated the training protocol of Fig.~\ref{fig:Applicaton}a three times to show the reproducibility of MRM in Fig.~\ref{fig:Applicaton}b.
One can design different training protocols at $T<30$~K in \ch{Eu5In2As6} to obtain different resistive states.
We hope that our results encourage the search for MRM in other magnetic semiconductors at even higher temperatures.
Such materials can be the centerpiece in device architectures for in-memory computations. \cite{liu_cryogenic_2025}

%%%%%%%%%%%%%%%%%%%%%%%%%%%%%%%%%%
%%%%%%%%%%%%% METHODS %%%%%%%%%%%%
%%%%%%%%%%%%%%%%%%%%%%%%%%%%%%%%%%
\pagebreak
\begin{methods}
\subsection{Material preparation.} Single crystals of \ch{Eu5In2As6} were grown via a Bi-flux method. 
High purity Eu, In, As, and Bi were mixed in the ratio 5:2:6:20, placed in an alumina crucible, and sealed under vacuum in a silica tube.
The tube was placed in a box furnace and heated to 950\C, held for 24 hours, then cooled to 700\C. 
After annealing for 12 hours, the samples were centrifuged to decant the flux. 
The crystals were shiny and needle-like, with the long dimension (2--5~mm) oriented along the $c$-axis.

\subsection{Electrical, thermodynamic, and optical characterizations.} Electrical resistivity and Hall effect were measured using four-probe techniques in a Quantum Design Dynacool-PPMS. 
The field was swept at 100~Oe/s, with stabilization at each field where the data were taken. 
No change in the behavior was observed with different field sweep rates. 
Magnetization data were obtained using a Quantum Design MPMS-3. 
A differential, membrane-based nanocalorimeter~\cite{tagliati_differential_2012} was used in a dilution refrigerator to measure heat capacity.
%Sanding of the samples is done under a fume hood to avoid arsenic gas poisoning. \sbc{We can remove this section if it's too obvious and add some detail in the above section}
A whisker-like single crystal of \ch{Eu5In2As6}, roughly $0.5\,{\rm mm}\times2\,{\rm mm}$, was mounted on the tip of an optically-black cone and the reflectivity determined over a wide frequency range with a Bruker Vertex 80v using an overfilling technique that employs {\it in situ} evaporation to establish the reference~\cite{homes1993}.

\subsection{X-ray diffraction.} Single-crystal diffraction was performed on a high-quality crystal in a Huber four-circle diffractometer in a closed-cycle ARM cryostat with base temperature of 4.5~K and resolution of 0.1~K. 
The beam-spot was $0.4\mathrm{mm}\times8\mathrm{mm}$ (horizonal$\times$vertical) for the needle-like samples with dimensions $3\mathrm{mm}\times0.5\mathrm{mm}$.
Data were collected using a Mythen 1K line detector with 640 channels and a $2\theta$ resolution of 0.005$^{\circ}$.
To ensure consistency of the scattering volume, the $\mathrm{(6 6 0)}$ and ${\mathrm(6 6 1)}$ Bragg peaks were studied individually, while reorienting the UB matrix at every temperature.

\subsection{Neutron diffraction.}
The neutron diffraction data were acquired with the thermal neutron triple-axis spectrometer VERITAS at the High Flux Isotope Reactor (HFIR) in Oak Ridge National Laboratory (ORNL). 
A single crystal of \ch{Eu5In2As6} was mounted in a 6 T cryomagnet with its $b$-axis parallel to the vertical field, and a base temperature of about 1.5~K. 
This geometry allows us to probe the (H,0,L) scattering plane using neutrons with a wavelength of 2.37~\AA. 
The statistical error for each intensity point corresponds to 1 standard deviation.

\end{methods}

%%%%%%%%%%%%%%%%%%%%%%%%%%%%%%%%%%
%%%%%%%%%%% REFERENCES %%%%%%%%%%%
%%%%%%%%%%%%%%%%%%%%%%%%%%%%%%%%%%

% Put the bibliography here, most people will use BiBTeX in
% which case the environment below should be replaced with
% the \bibliography{} command.

\pagebreak
\section*{References}
\bibliographystyle{naturemag} % choose style first
\bibliography{Bibliography}
%\nocite{*}

% %%%%%%%%%%%%%%%%%%%%%%%%%%%%%%%%%%
% %%%%%%%%%%%% ADDENDUM %%%%%%%%%%%%
% %%%%%%%%%%%%%%%%%%%%%%%%%%%%%%%%%%

%% Here is the endmatter stuff: Supplementary Info, etc.
%% Use \item's to separate, default label is "Acknowledgements"
\pagebreak
\begin{addendum}
 \item[Acknowledgments]
 We thank Itamar Kimchi and Rafael Fernandes for insightful discussions. 
 The work at Boston College (transport and thermodynamic measurements) was funded by the U.S. Department of Energy, Office of Basic Energy Sciences, Division of Physical Behavior of Materials under award number DE-SC0023124.
The neutron scattering experiments by S.B. and E.O.G.D. were supported by the National Science Foundation under award number DMREF-2522383.
 This material is based upon work supported by the Air Force Office of Scientific Research under award number FA9550-23-1-0124.
 Work at UCSD was supported by the National Science Foundation under Grant No. DMR-2145080.
 A portion of this work was performed at the National High Magnetic Field Laboratory, which is supported by the National Science Foundation Cooperative Agreement No. DMR-1644779 and the state of Florida. 
 A portion of this research used resources at the High Flux Isotope Reactor, a DOE Office of Science User Facility operated by the Oak Ridge National Laboratory. 
The beamtime was allocated to VERITAS on proposal number IPTS-35500.
The support for neutron scattering was provided by the Center for High-Resolution Neutron Scattering, a partnership between the National Institute of Standards and Technology and the National Science Foundation under Agreement No. DMR-2010792. 
The identification of any commercial product or trade name does not imply endorsement or recommendation by the National Institute of Standards and Technology.

 \item[Author Contributions]  
 S.R.B. performed transport and thermodynamic measurements, analyzed data, and wrote the first draft of the manuscript.  
 M.B.M. and S.R.B. grew crystals and characterized them.
 R.B. and A.F. performed high-resolution X-ray measurements.
 E.O.G.D., A.A.A, and J.G. performed neutron diffraction.
 A.R. performed nanocalorimetric measurements.
 D.E.G. assisted with high-field experiments.
 C.C.H. performed optical conductivity experiments.
 Y.R. provided theoretical input. 
 F.T. conceptualized and coordinated the project. 
 All coauthors participated in the writing process.  
 \item[Inclusion and Ethics] Contributions from all authors, including local scientists at the national labs, are properly acknowledged in this work.
 \item[Competing interests] The authors declare no competing interests.
 \item[Data availability] All data in this work will be made available online after publication.
 \item[Supplementary information] A PDF file containing supplementary transport, magnetization, specific heat, x-ray, and optical reflectivity data is available online.
 \item[Correspondence and requests for materials] should be addressed to F.T. via email: fazel.tafti@bc.edu

%%%%%%%%%%%%%%%%%%%%%%%%%%%%%%%%%%%%%%%%%%%%%%%%
%\begin{document}
%\preprint{APS/123-QED}


%\pacs{71.20.Eh, ?????}% PACS, the Physics and Astronomy

%\tableofcontents

\begin{center}
{\large \textbf{Supplementary Information}}\\[6pt]
{\large \textbf{Magnetoresistive Memory in the Paramagnetic Phase of \ch{Eu5In2As6}}}
\end{center}

\setcounter{figure}{0}
\renewcommand{\thefigure}{S\arabic{figure}}

\setcounter{table}{0}
\renewcommand{\thetable}{S\arabic{table}}

\section{\label{sec:gap}Semiconducting Properties}
%%%%%%%%%%%%%%%%%% FIGURE1 %%%%%%%%%%%%%%%%%%%
\begin{figure}
  \includegraphics[width=0.46\textwidth]{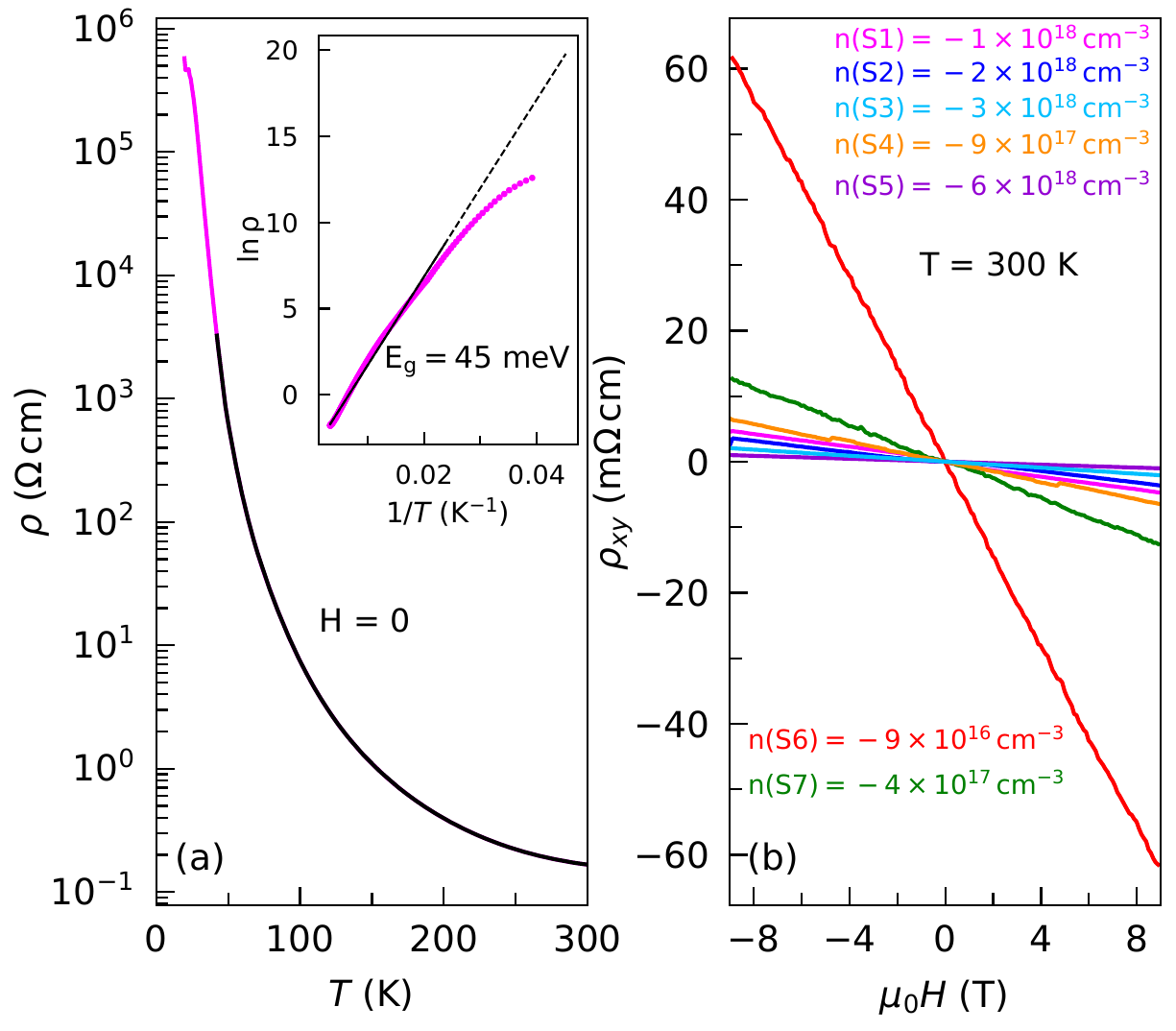}
  \caption{\label{fig:Carriers}
  (a) Zero-field resistivity of the sample in the main text, from which a gap of 45(6)~meV is extracted by fitting the data to a single exponential (black line).
  (b) Hall resistivity plotted as a function of field in 7 samples, showing $n$-type carriers with concentrations between $10^{16}$ and $10^{18}$~cm$^{-3}$. 
  }
\end{figure}
%%%%%%%%%%%%%%%%%%%%%%%%%%%%%%%%%%%%%%%%%%%%%%
\ch{Eu5In2As6} is a magnetic semiconductor with a small gap of 45~meV extracted from the Arrhenius analysis in Fig.~\ref{fig:Carriers}a.
The data were obtained from the same sample used in the main text (S1).
Reflectivity measurements in Figs.~\ref{fig:reflec} and \ref{fig:sigma} show an optical gap of 60~meV, close to the transport gap.

Hall effect was measured on several samples (S1 through S7) at room temperature, all showing a negative slope that indicates $n$-type carriers (Fig.~\ref{fig:Carriers}b).
Using a single band model, $R_H=\rho_{xy}/H=1/ne$, the carrier concentration was evaluated and reported on Fig.~\ref{fig:Carriers}b.
Since \ch{Eu5In2As6} is a self-doped semiconductor, the carrier concentration could vary among different samples due to a slight off-stoichiometry.
However, it is always between $10^{16}$ and $10^{18}$~cm$^{-3}$, at least four orders of magnitude smaller than those of manganites and iridates in their metallic phases.
Remarkably, such a small concentration of extrinsic carriers produces extremely large CMR and MRM effects. 

We measured the resistivity of a \ch{Eu5In2As6} sample at 300~mK as a function of the magnetic field up to 40~T, and did not find quantum oscillations, confirming the absence of a metallic ground state.
Plugging the room-temperature carrier density $n\approx10^{18}$~cm$^{-3}$ and resistivity $\rho\approx0.1~\Omega\,$cm (Fig.~\ref{fig:Carriers}) into the Drude expression $\rho=1/ne\mu$ yields a small mobility $\mu\approx 60$~cm$^2$V$^{-1}$s$^{-1}$, consistent with the absence of quantum oscillations and indicating a hopping-like motion of charge carriers.

\section{\label{sec:gap}Sample dependence of MRM}
%%%%%%%%%%%%%%%%%% FIGURE2 %%%%%%%%%%%%%%%%%%%
\begin{figure}
  \includegraphics[width=0.46\textwidth]{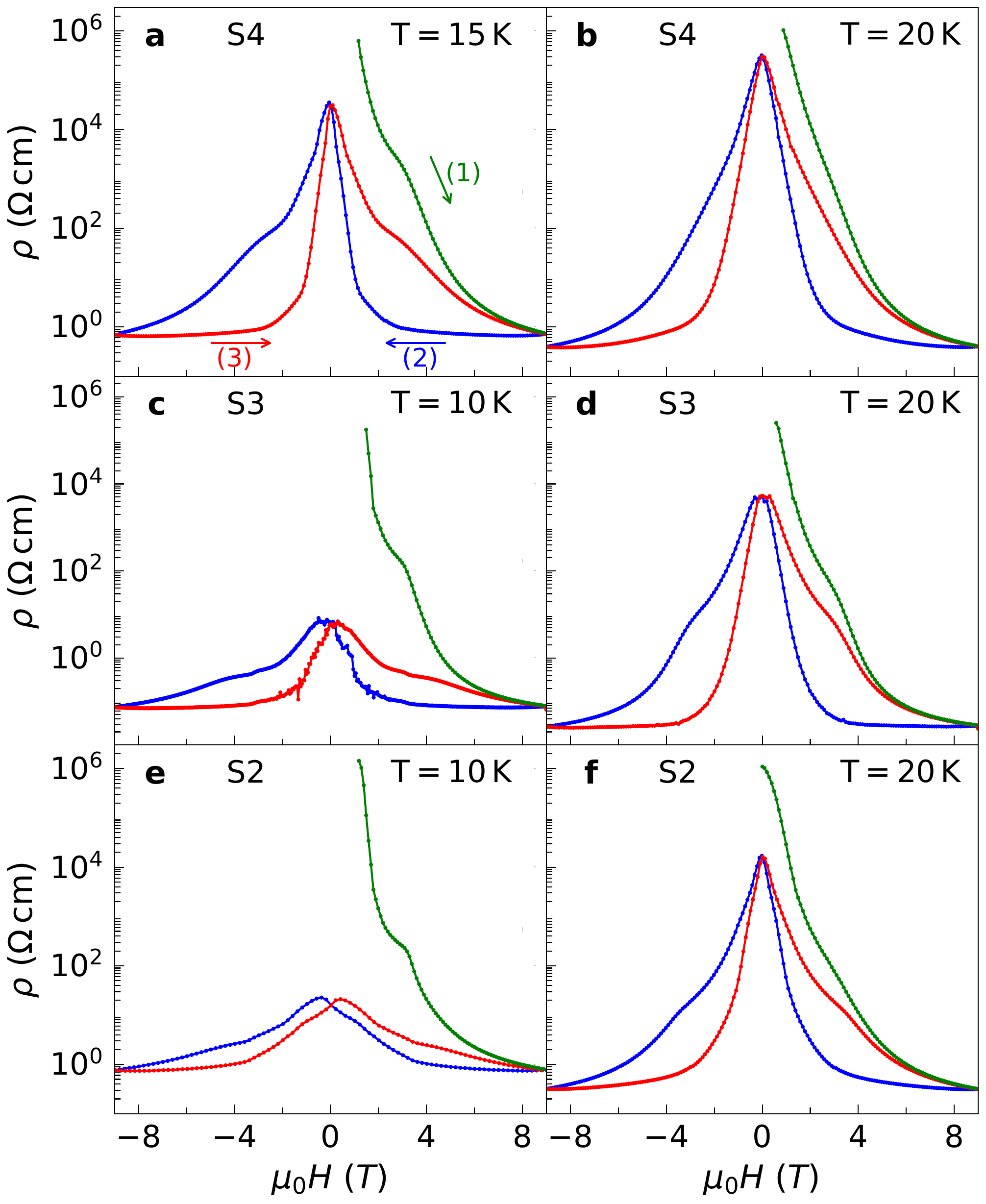}
  \caption{\label{fig:RH_SM}
  Resistivity as a function of field at both $T<T_\text{N}$ (left column) and $T>T_\text{N}$ (right column) in three different \ch{Eu5In2As6} samples.
  The general features of MRM, observed in sample S1 (main text), are reproduced in other samples. 
  }
\end{figure}
%%%%%%%%%%%%%%%%%%%%%%%%%%%%%%%%%%%%%%%%%%%%%%
In Fig.~\ref{fig:RH_SM}, we show the reproducibility of MRM among different samples. 
Both characteristic features of MRM in $\rho(H)$, namely the difference between untrained and trained resistivity values near zero-field and the hysteretic loop between upsweep and downsweep data, are reproduced in different samples (Fig.~\ref{fig:RH_SM}).
%These samples are different from S1 used in the main text.

%%%%%%%%%%%%%%%%%% Table1 %%%%%%%%%%%%%%%%%%%
\begin{table}[b]
\caption{Sample dependence of MRM.}
\label{tab:lattice}

\begin{tabular}{ccc}
Sample & $n$~(cm$^{-3}$) & MRM \\
\hline
S1 & $1\times 10^{18}$ & Yes \\
S2 & $2\times 10^{18}$ & Yes \\
S3 & $3\times 10^{18}$ & Yes \\
S4 & $9\times 10^{17}$ & Yes \\
S5 & $6\times 10^{18}$ & Yes \\
S6 & $9\times 10^{16}$ & No \\
S7 & $4\times 10^{17}$ & No \\
\end{tabular}

\end{table}
%%%%%%%%%%%%%%%%%%%%%%%%%%%%%%%%%%%%%%%%%%%%%

All \ch{Eu5In2As6} samples exhibit CMR, with decreasing resistivity values as the magnetic field is increased.
However, the resistivity of samples with carrier concentrations less than $5 \times 10^{17}$~cm$^{-3}$ remains insulating even at 9~T.
Such samples do not exhibit MRM.
For example, among the samples shown in Fig.~\ref{fig:Carriers}b, S1, S2, S3, S4, and S5 exhibit MRM but S6 and S7 do not. 
This observation confirms that MRM is a property of a small number of extrinsic carriers, and if their concentration is below a certain threshold, the material will not show a memory effect.
We summarize the samples, their carrier concentrations, and whether they exhibit MRM in Table 1.

\section{\label{sec:Angle}Angle dependence of MRM}
%%%%%%%%%%%%%%%%%% FIGURE3 %%%%%%%%%%%%%%%%%%%
\begin{figure}
  \includegraphics[width=0.46\textwidth]{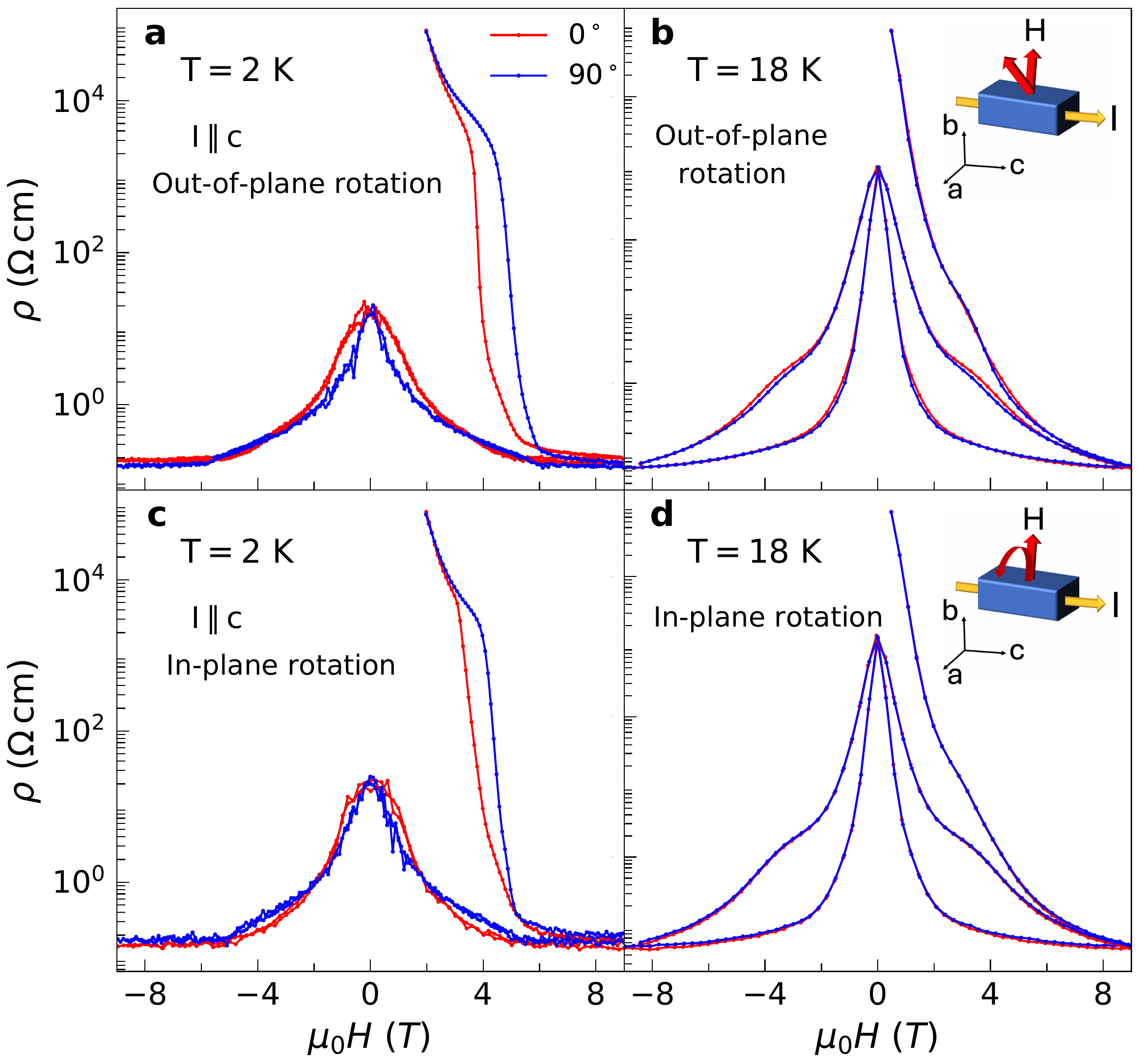}
  \caption{\label{fig:ANGLE}
  (a) Resistivity as a function of magnetic field at 2~K, with the current applied along the $c$-axis and the magnetic field rotated out of the $ab$ plane. 
  The angles listed in the legend are between the field and crystal's $c$-axis.
  (b) Same measurement as in (a) but at 18~K. 
  (c) Measurements while the field is rotated in the $ab$ plane at 2~K. 
  (d) Same measurement as in (c) but at 18~K.} 
  
\end{figure}
%%%%%%%%%%%%%%%%%%%%%%%%%%%%%%%%%%%%%%%%%%%%%%

\ch{Eu5In2As6} displays strong anisotropy in magnetization when the magnetic field points along different directions.
However, the resistivity remains qualitatively the same under different directions of the applied field, as shown in a prior publication [23]. 
We show the field dependence of resistivity as the magnetic field is rotated out of the $ab$-plane in Fig.~\ref{fig:ANGLE}a and b at 2~K and 18~K, respectively.
We also show the field dependence of resistivity as the magnetic field is rotated within the $ab$ plane in Fig.~\ref{fig:ANGLE}c and d at 2~K and 18~K, respectively. 
The MRM effect shows little dependence on the magnetic field orientation.

\section{\label{sec:CMR}Temperature dependence of CMR and MRM}
%%%%%%%%%%%%%%%%%% FIGURE4 %%%%%%%%%%%%%%%%%%%
\begin{figure*}
  \includegraphics[width=\textwidth]{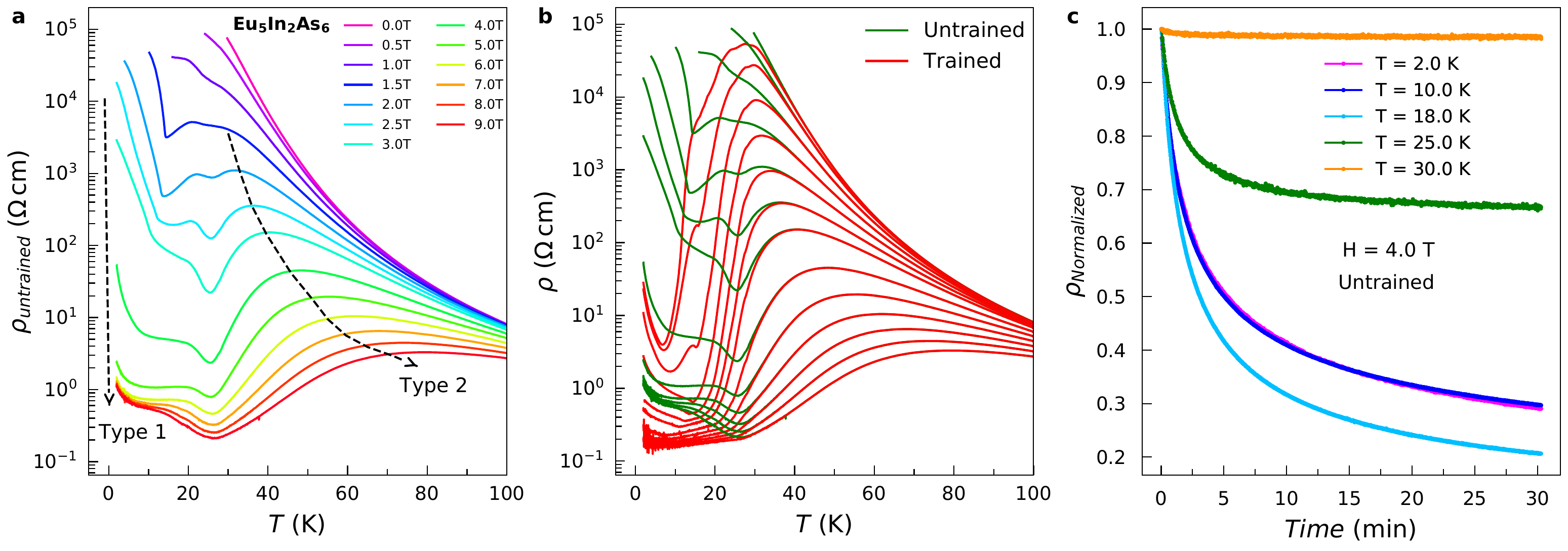}
  \caption{\label{fig:CMR}
  (a) Untrained resistivity as a function of temperature at several fields, revealing two types of CMR and a resistivity minimum. 
  (b) Untrained (green) and trained (red) resistivity at the same measuring fields as in panel (a). 
  The $\rho_\text{untrained}/\rho_\text{trained}$ is calculated from these data.
  MRM onsets near 30~K where the green and red curves split.
  (c) Normalized resistivity plotted as a function of temperature. 
  The relaxor-like behavior vanishes along with MRM above 30~K.} %as the hidden order disappears above 30~K.
  
\end{figure*}
%%%%%%%%%%%%%%%%%%%%%%%%%%%%%%%%%%%%%%%%%%%%%%
The CMR properties of \ch{Eu5In2As6} have been reported previously [23], but for completeness, the CMR effect of sample S1 (used in the main text) is shown in Fig.~\ref{fig:CMR}a. 
The untrained $\rho(T)$ curves are measured from 2 to 100~K under different measuring fields.
At low fields, the resistance is so hight that it cannot be measured reliably in the PPMS.
The CMR effect in Fig.~\ref{fig:CMR}a is a negative magnetoresistance reflected in the suppression of $\rho_\text{untrained}(T)$ curves with increasing magnetic field.
Reference [23] explains in detail that \ch{Eu5In2As6} shows two types of CMR, one at $T>T_\text{N}$ (suppression of the broad peak by field) and another at $T<T_\text{N}$ (suppression of the low-temperature upturn by field).

Figure~\ref{fig:CMR}b shows both $\rho_\text{untrained}(T)$ and $\rho_\text{trained}(T)$ curves in sample S1.
The training field is 9~T for all the red curves, and the measuring fields are the same as indicated in Fig.~\ref{fig:CMR}a.
The hidden order parameter $\rho_\text{untrained}/\rho_\text{trained}$ in the main text Fig.~2b is obtained by dividing the green curves by the red curves in Fig.~\ref{fig:CMR}b.

Note that MRM (the separation between red and green curves in Fig.~\ref{fig:CMR}b) onsets near 30~K, slightly higher than the temperature of the resistivity minimum in untrained data ($T_\text{min}$). %untrained resistivity shows a minimum.
Below 30~K, the untrained resistivity shows a relaxor-like behavior, as discussed in the main Fig.~3.
As shown in Fig.~\ref{fig:CMR}c, the relaxor-like behavior disappears above 30~K.
%In summary,  Fig.~\ref{fig:CMR} shows a resistivity minimum in the untrained resistivity near 30~K, below which, both MRM and resistivity relaxation appear.
In summary,  Fig.~\ref{fig:CMR} shows that the MRM, resistivity minimum in untrained curves, and resistivity relaxation all appear below 30~K.

\section{\label{sec:Comp}Comparison with the antimony version}
%%%%%%%%%%%%%%%%%% FIGURE5 %%%%%%%%%%%%%%%%%%%
\begin{figure}
  \includegraphics[width=0.46\textwidth]{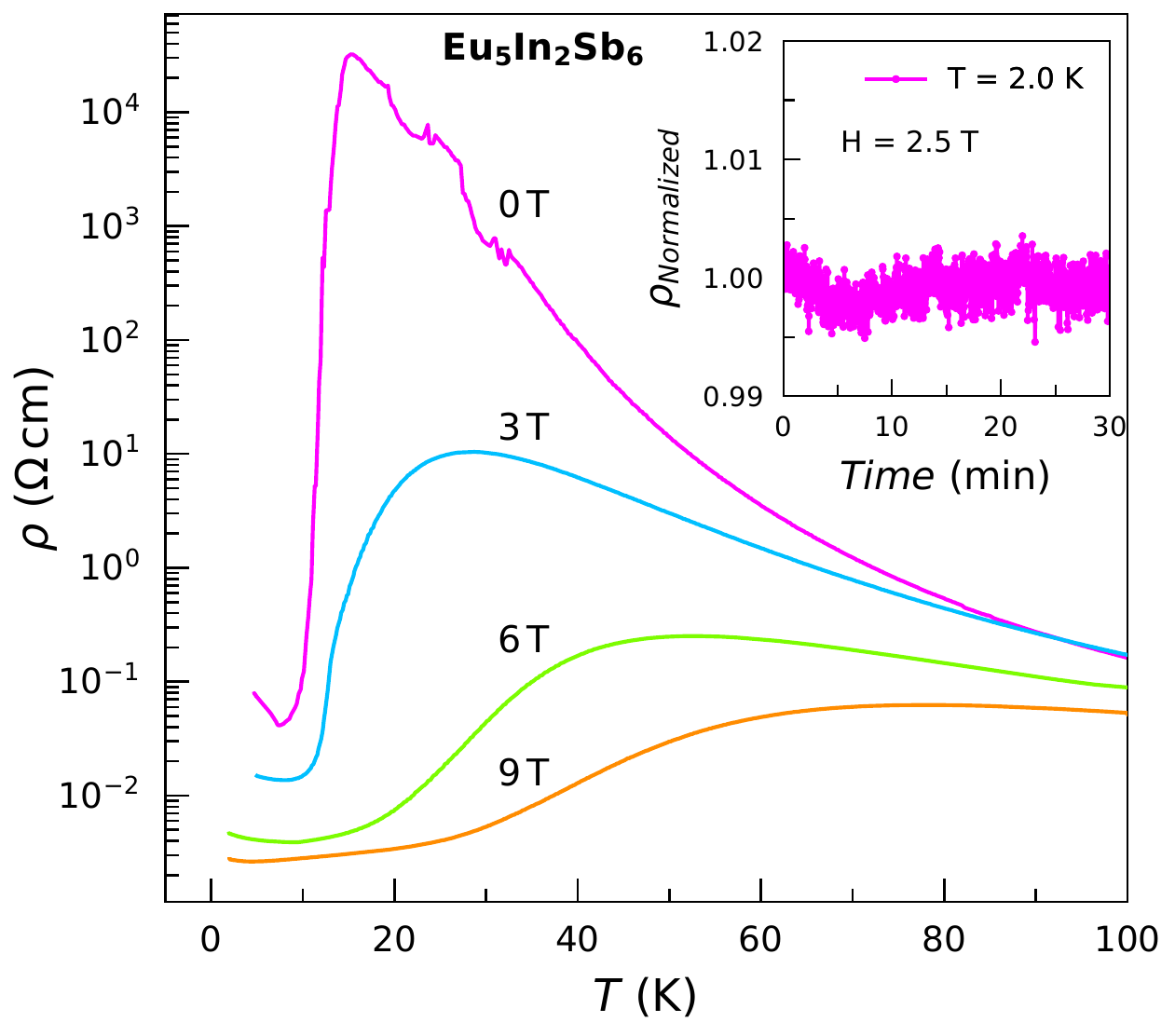}
  \caption{\label{fig:Eu526}
Resistivity as a function of temperature in \ch{Eu5In2Sb6}. 
There is no resistivity minimum at zero and other fields, and we did not find MRM in this compound.
The inset shows time-independent resistivity.
}
\end{figure}
%%%%%%%%%%%%%%%%%%%%%%%%%%%%%%%%%%%%%%%%%%%%%%%%
We did not observe either MRM or time-dependence of resistivity in \ch{Eu5In2Sb6}, a sister compound of \ch{Eu5In2As6}. 
A key distinction is the absence of a resistivity minimum in \ch{Eu5In2Sb6} (Fig.~\ref{fig:Eu526}), unlike in \ch{Eu5In2As6} where MRM onsets slightly above $T_\text{min}$, the temperature of resistivity minimum (Fig.~\ref{fig:CMR}). %with a resistivity minimum near 30~K (Fig.~\ref{fig:CMR}).
The comparison between the two sister compounds suggests that MRM relies on the hybridization between $f$-moments and conduction electrons, which is reflected in the resistivity minimum.
The antimony version shows no resistivity minimum and is devoid of MRM.

\section{\label{sec:AC}AC susceptibility}
%%%%%%%%%%%%%%%%%% FIGURE6 %%%%%%%%%%%%%%%%%%%
\begin{figure}
  \includegraphics[width=0.46\textwidth]{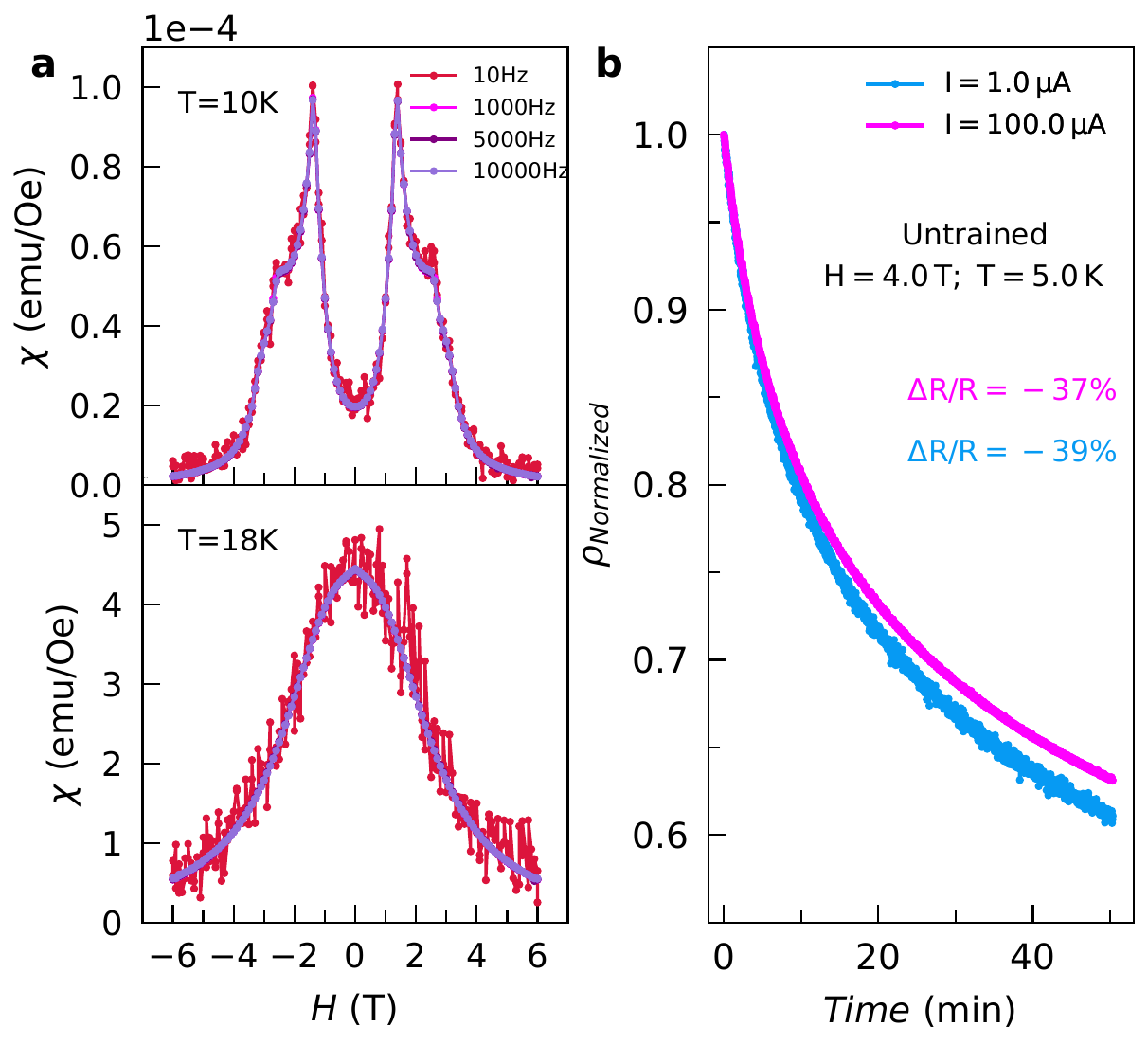}
  \caption{\label{fig:AC}
  (a) AC susceptibility measured in \ch{Eu5In2As6} at $T<T_\text{N}$ and $T>T_\text{N}$ using several frequencies, showing no frequency dependence.
  (b) Time dependence of resistivity, measured at $H=4$~T under a ZFC condition. Increasing the current by two orders of magnitude has little impact on the resistivity relaxation. 
  }
\end{figure}
%%%%%%%%%%%%%%%%%%%%%%%%%%%%%%%%%%%%%%%%%%%%%%
The AC susceptibility of \ch{Eu5In2As6} is independent of the frequency of the applied field (Fig.~\ref{fig:AC}).
This is true for both $T<T_\text{N}$ and $T>T_\text{N}$.
Thus, we rule out inhomogeneous magnetism and spin-glass behavior as the mechanism of MRM.

To rule out Joule heating effects, we show resistivity data as a function of time, measured with two different current values in Fig.~\ref{fig:AC}b.
The relaxation behavior is not changed considerably, despite increasing the current by two orders of magnitude from $1~\mu\mathrm{A}$ to $100~\mu\mathrm{A}$, hence increasing the Joule heating by four orders of magnitude.
This demonstrates that the resistivity relaxation is intrinsic; It is not an artifact of Joule heating.

\section{\label{sec:Xray}X-ray Diffraction Analysis}
%%%%%%%%%%%%%%%%%% FIGURE7 %%%%%%%%%%%%%%%%%%%
\begin{figure}
  \includegraphics[width=0.46\textwidth]{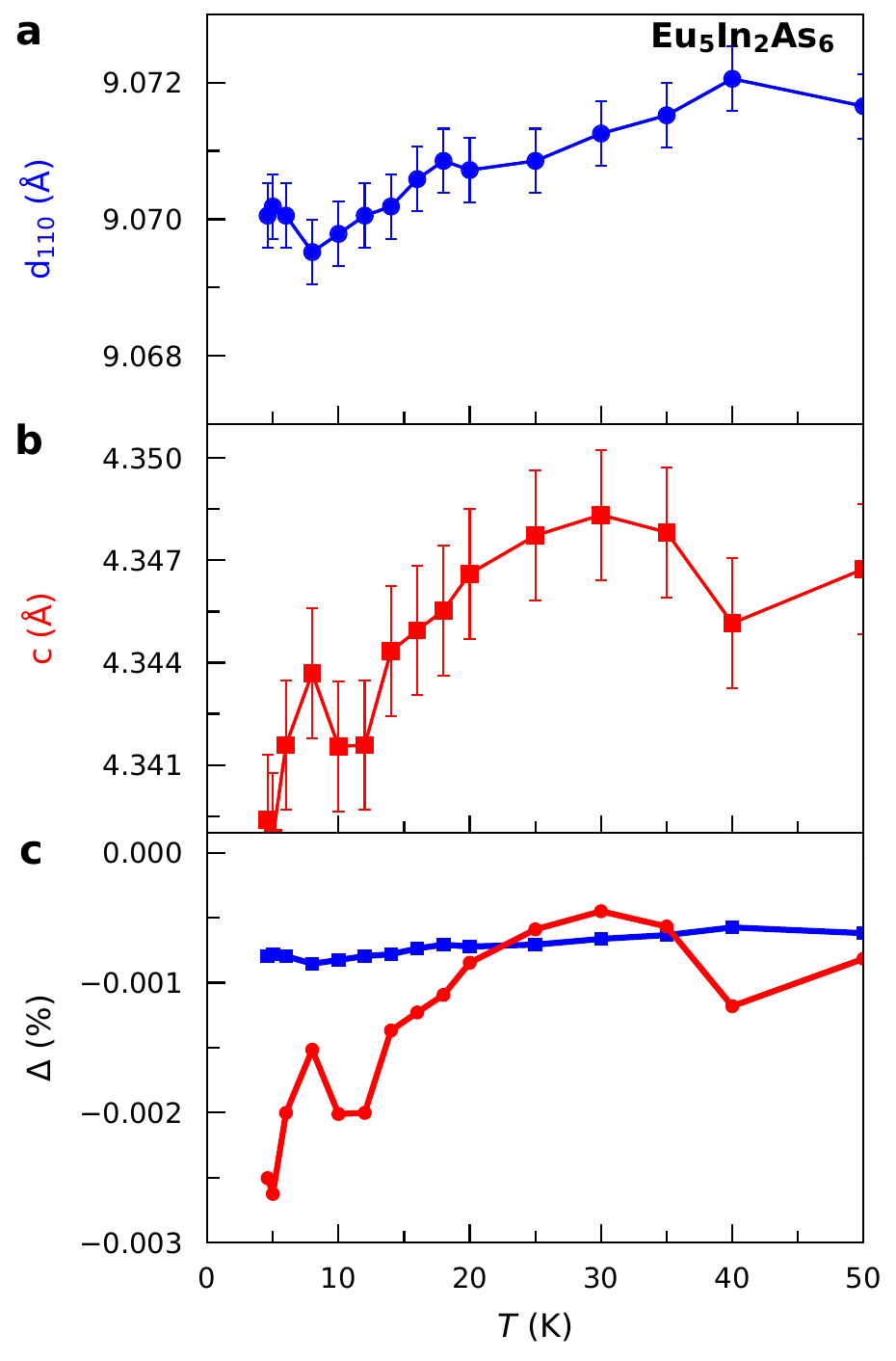}
  \caption{\label{fig:Xray}
  The change of lattice dimensions along two orthogonal directions (a) [110], and (b) [001] as a function of temperature in \ch{Eu5In2As6} from x-ray diffraction data shown between 5~K and 50~K. Errors are propagated from the uncertainty in the angular measure of 2$\theta$. (c) Percentage change with respect to the corresponding 100~K value depicts the anisotropy of the lattice change.}
  
\end{figure}
%%%%%%%%%%%%%%%%%%%%%%%%%%%%%%%%%%%%%%%%%%%%%%%%

%%%%%%%%%%%%%%%%%% Table1 %%%%%%%%%%%%%%%%%%%
\begin{table}[t]
\caption{Temperature dependence of lattice parameters.}
\label{tab:lattice}

\begin{tabular}{cccc}
$T$ (K) & $d_{110}$ (\AA) & $c$ (\AA) \\
\hline
100.0 & 9.07728 $\pm$ 0.00011 & 4.35029 $\pm$ 0.00382 \\
75.0  & 9.077473 $\pm$ 0.00011 & 4.34499 $\pm$ 0.00381 \\
50.0  & 9.077166 $\pm$ 0.00011 & 4.34674 $\pm$ 0.00381 \\
40.0  & 9.07206 $\pm$ 0.00011 & 4.34515 $\pm$ 0.00381 \\
35.0  & 9.07152 $\pm$ 0.00011 & 4.34782 $\pm$ 0.00382 \\
30.0  & 9.07125 $\pm$ 0.00011 & 4.34833 $\pm$ 0.00382 \\
25.0  & 9.07085 $\pm$ 0.00011 & 4.34773 $\pm$ 0.00382 \\
20.0  & 9.07072 $\pm$ 0.00011 & 4.34661 $\pm$ 0.00381 \\
18.0  & 9.07085 $\pm$ 0.00011 & 4.34553 $\pm$ 0.00381 \\
16.0  & 9.07059 $\pm$ 0.00011 & 4.34494 $\pm$ 0.00381 \\
14.0  & 9.07019 $\pm$ 0.00011 & 4.34434 $\pm$ 0.00381 \\
12.0  & 9.07005 $\pm$ 0.00011 & 4.34158 $\pm$ 0.00380 \\
10.0  & 9.06978 $\pm$ 0.00011 & 4.34155 $\pm$ 0.00380 \\
8.0   & 9.06952 $\pm$ 0.00011 & 4.34370 $\pm$ 0.00381 \\
6.0   & 9.07005 $\pm$ 0.00011 & 4.34158 $\pm$ 0.00380 \\
5.0   & 9.07019 $\pm$ 0.00011 & 4.33887 $\pm$ 0.00379 \\
4.6   & 9.07005 $\pm$ 0.00011 & 4.33940 $\pm$ 0.00379 \\
\end{tabular}

\end{table}
%%%%%%%%%%%%%%%%%%%%%%%%%%%%%%%%%%%%%%%%%%%%%
Figure~\ref{fig:Xray} presents the temperature dependence of the lattice parameters, determined from the measured $2\theta$ values as the sample was gradually warmed from the base temperature to 50~K. 
%A noticeable contraction of the lattice parameter along the $\mathrm{[1,1,0]}$ direction sets in below approximately 35~K. 
%The behavior of the out-of-plane parameter, $c$, is even more striking. 
The in-plane and out-of-plane lattice parameters were extracted from the (660) and (661) Bragg peaks.
From the (660) Bragg peak, we extract the $d$-spacing parallel to (110) direction following the Bragg condition:

\begin{equation*}
    \lambda_\mathrm{Cu-K\alpha} = 2d_{HKL}\times \sin(\theta) = 2 \frac{1}{\sqrt{\frac{H^2}{a^2}+\frac{K^2}{b^2}+\frac{L^2}{c^2}}}\times \sin(\theta)
\end{equation*}

Taking the value for the $\lambda_\mathrm{Cu-K\alpha}$ to be $1.54$ \AA, we found the value of $d_{110} = \frac{ab}{\sqrt{a^2+b^2}}$ which is listed in Table~1 and plotted in Fig.~\ref{fig:Xray}. 
The $c$ lattice parameter is extracted using the same equation and known values of $d_{110}$ while measuring the Bragg angle of reflection of the (661) peak. 
Since these are derived quantities, we propagate the error that influences the $c$ lattice parameter the most.

Although the lattice parameters are derived and thus subject to a larger uncertainty than the peak positions shown in the main text, their evolution nonetheless reveals a clear trend. 
Excluding the anomalous measurement at 40~K, the $c$ lattice parameter contracts below 30~K, where MRM appears.
The rate of contraction increases below the AFM ordering temperature at $T_\text{N}=16$~K. 
The thermal contraction appears to halt near 10~K, followed by a slight expansion before resuming again below the spin-reorientation transition at $T_{\mathrm{N}}^\prime \approx 6~\text{K}$. 

%%%%%%%%%%%%%%%%%% FIGURE8 %%%%%%%%%%%%%%%%%%%
\begin{figure*}
  \includegraphics[width=\textwidth]{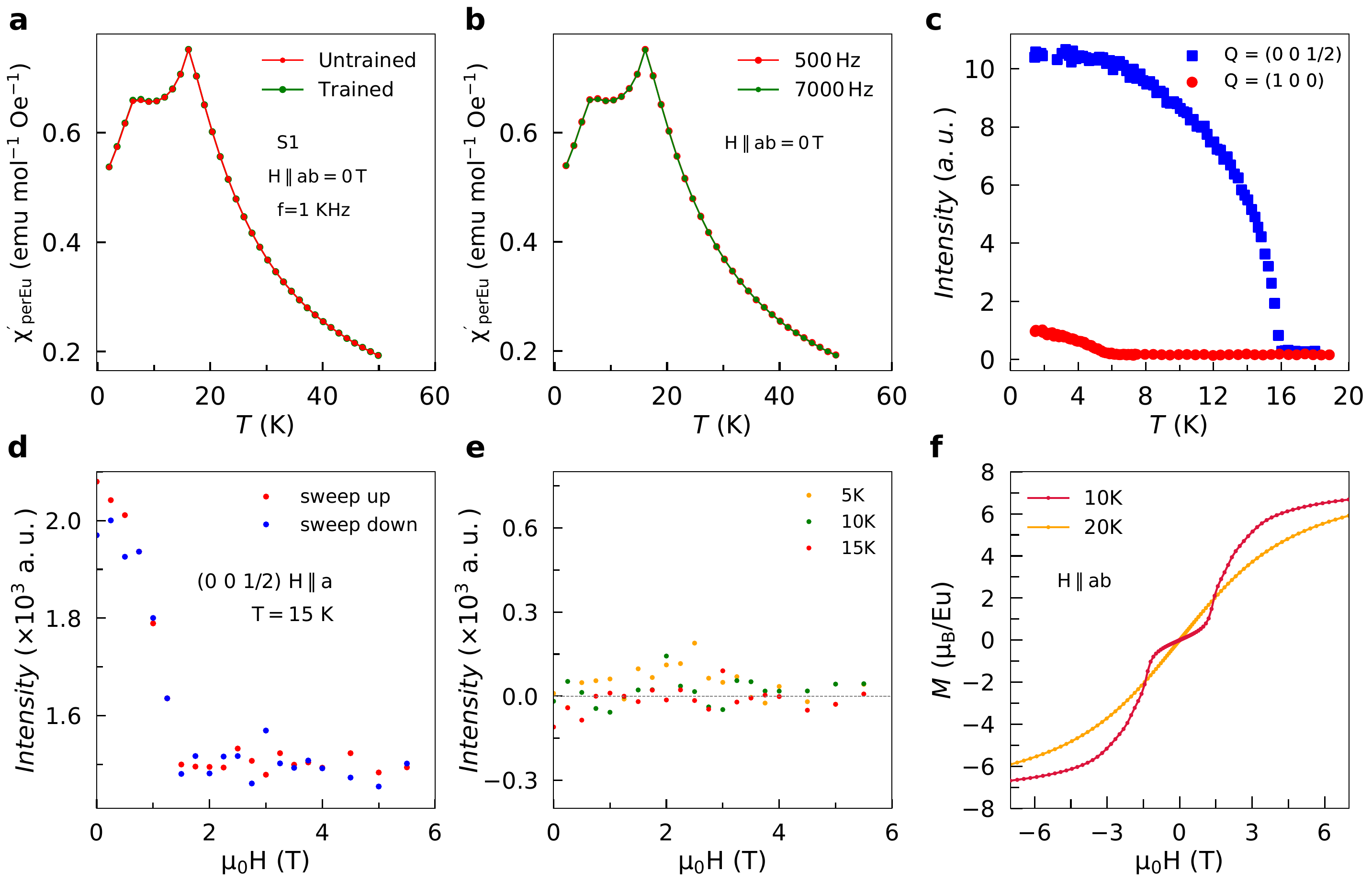}
  \caption{\label{fig:LRO}
  (a) Susceptibility as a function of temperature in an untrained (ZFC) state and after training the sample under a 9~T field.
  (b) Susceptibility as a function of temperature measured at different frequencies.
  (c) Neutron scattering intensity as a function of temperature reveals two AFM transitions at 16~K and 6~K.
  (d) Neutron scattering intensity as a function of magnetic field during up and down field sweeps shows no hysteresis.
  (e) Difference between the up and down sweep curves.
  (f) Magnetization as a function of magnetic field at 10~K and 20~K.
  }
\end{figure*}
%%%%%%%%%%%%%%%%%%%%%%%%%%%%%%%%%%%%%%%%%%%%%%
Since the lattice spacing related to the Bragg peak (661) contains contributions from all three lattice parameters, it is sensitive to a change in any of them. 
A question can be asked if the lattice parameters show any anisotropy in their behavior. 
In order to answer that, we depicted the percentage change ($\Delta$) in each lattice parameter with respect to the corresponding 100~K value in Fig.~\ref{fig:Xray}c. 
The change along $d_{110}$ records the out-of-plane lattice spacing, while along $c$ records the in-plane lattice spacing. 
We observe that the primary contraction happens along the $c$-lattice parameter, which is along one of the in-plane directions of the needle-like crystal. 
Further complementary measurements using a dilatometer should follow, confirming the axis of contraction.

\section{\label{sec:neutron}Absence of Long-Range Order above 16~K}
\ch{Eu5In2As6} does not support any long-range magnetic order at 30~K, where MRM starts.
Its AFM order does not develop until 16~K, nearly half the onset temperature of MRM.

Here, we present supplementary data confirming the absence of a long-range order in \ch{Eu5In2As6} above 16~K on the same sample (S1) used in the main manuscript.
Figure~\ref{fig:LRO}a shows the susceptibility as a function of temperature under untrained and trained (9~T) conditions.
There is no difference between the trained and untrained susceptibility curves, in stark contrast to the transport data. 
The AC susceptibility is also independent of drive frequencies (Fig.~\ref{fig:LRO}b), which rules out a spin glass phase in the title compound. 
Figure~\ref{fig:LRO}c shows our neutron diffraction data, revealing a long-range $Q=(0\,0\,\frac{1}{2})$ ordering wave vector below 16~K and another $Q=(1\,0\,0)$ wave vector below 6~K.
We did not find any other magnetic wave vectors, so there are no FM or AFM orders at $T>16$~K.

Within the AFM ordered state, we found no hysteresis in the $Q=(0\,0\,\frac{1}{2})$ wave vector diffraction data between up and down field sweeps, unlike the resistivity hysteresis loops.
The absence of such a hysteresis in the diffraction data is better resolved in Fig.~\ref{fig:LRO}e, which traces the difference between the two curves in Fig.~\ref{fig:LRO}d.

Since there is no long-range order at $16<T<30$~K, MRM must result from either a hidden order or short-range correlations (polarons). 
Consistent with this interpretation, Fig.~\ref{fig:LRO}e shows the absence of magnetization hysteresis at 10~K and 20~K, below and above $T_\text{N}$. 
This is in contrast to manganites which show hysteretic behavior under field sweeps in both magnetization and neutron scattering measurements[3,5,6].

\section{\label{sec:optical}Optical Conductivity}
%%%%%%%%%%%%%%%%%% FIGURE7 %%%%%%%%%%%%%%%%%%%
\begin{figure}
  \includegraphics[width=0.46\textwidth]{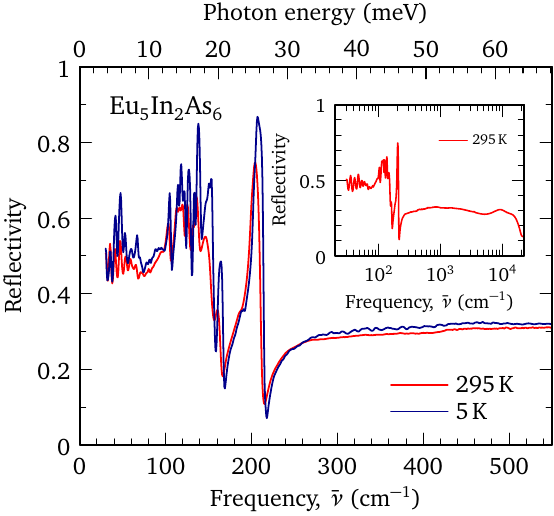}
  \caption{\label{fig:reflec}
  The reflectivity of a single crystal of Eu$_5$In$_2$As$_6$ versus wave number in the far-infrared region for light polarized in the \emph{a-b} planes at 5 and 295~K, where the unscreened infrared-active lattice vibrations dominate the response. The resolution at low frequency is typically better than 2~cm$^{-1}$. Inset: semilog plot of the reflectivity at room temperature over a wide frequency range. 
  }
\end{figure}
%%%%%%%%%%%%%%%%%%%%%%%%%%%%%%%%%%%%%%%%%%%%%%
We performed complementary optical conductivity measurements at zero field to confirm the semiconducting nature of \ch{Eu5In2As6} and rule out the presence of an underlying Fermi surface.
Temperature dependence of reflectivity was measured in the far- and mid-infrared regions, but little temperature dependence was observed above 50~meV (Fig.~\ref{fig:reflec}).  
The behavior can be described as semiconducting, where the low-frequency region is dominated by the normally infrared-active lattice modes, and the high-frequency region (mid-infrared and above) is relatively flat, decreasing slightly above $\sim 2$~eV.
The interference fringes at low frequencies result from back reflections due to the slab-like geometry of the sample and the absence of free-carrier absorption.

The large number of overlapping phonon line shapes complicates the response and makes interpretation difficult.  
It is easier to examine the imaginary part of the dielectric function (absorption), or the real part of the optical conductivity.  
To this end, the complex dielectric function, $\tilde\epsilon(\omega) = \epsilon_1 + i\epsilon_2$, has been determined from a Kramers-Kronig analysis of the reflectivity, 
which requires extrapolations at high and low frequencies.   
At low frequency, the reflectivity is assumed to be constant below the lowest measured frequency, $R(\omega\rightarrow 0) \simeq 0.5$.
Above the highest-measured frequency, the reflectance was assumed to follow a $\omega^{-1}$ dependence up to $15\times 10^4$~cm$^{-1}$, above which a free-electron approximation ($R\propto \omega^{-4}$) was assumed.  
The real part of the optical conductivity is defined as:
\begin{equation}
  \sigma_1(\omega)= \left(\frac{2\pi}{Z_0}\right)\,\omega\epsilon_2  \ \ (\Omega^{-1}{\rm cm}^{-1})
\end{equation}
where $Z_0\simeq 377$~$\Omega$ is the impedance of free space. 

The optical conductivity is shown over a wide range in Fig.~\ref{fig:sigma}, while the inset displays a narrower frequency interval in the region of the infrared-active modes.  The optical conductivity shows an onset of absorption above about 500 cm$^{-1}$ ($\sim 60$~meV), which is identified as the optical gap.  
The optical gap is somewhat larger than the transport gap of $\sim 45$~meV.
This is because the optical gap probes only direct transitions, while the transport gap probes indirect (phonon-assisted) transitions that typically yield a lower gap.

%%%%%%%%%%%%%%%%%% FIGURE8 %%%%%%%%%%%%%%%%%%%
\begin{figure}
  \includegraphics[width=0.46\textwidth]{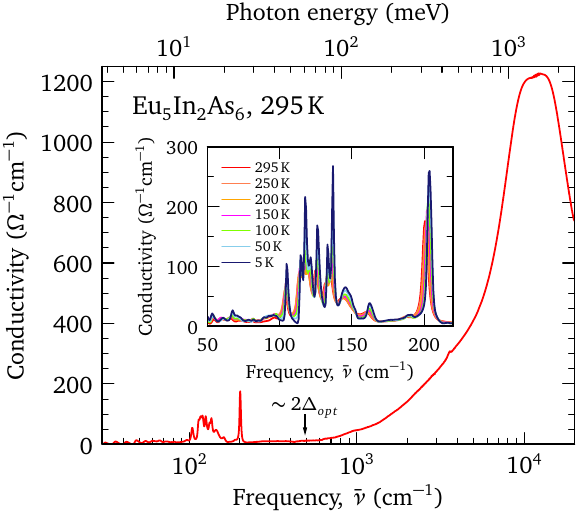}
  \caption{\label{fig:sigma}
  The optical conductivity of Eu$_5$In$_2$As$_6$ at room temperature over a wide frequency range showing the onset of absorption due to direct transitions above about $2\Delta_{opt}\sim 500$~cm$^{-1}$.  The absorption due to the lattice modes is mostly confined to the narrow interval between 100 and 200~cm$^{-1}$. 
  Inset: the optical conductivity in the low-frequency region of the normally infrared-active
  vibrations.
  }
\end{figure}
%%%%%%%%%%%%%%%%%%%%%%%%%%%%%%%%%%%%%%%%%%%%%%

\section{\label{sec:theory}Theoretical Discussion}
Now we discuss the possible symmetry origin of the domain-induced hysteresis observed in magnetoresistance. 
The issue is to identify the symmetry of the hidden order parameter. 
The point group of \ch{Eu5In2As6} is $D_{2h}$, featuring 8 distinct 1D irreducible representations (irreps). 
The magnetic field $\vec H$ pseudovector transforms as $B_{1g}$, $B_{2g}$, and $B_{3g}$ irreps for the $z$, $y$ and $x$ component, respectively. 
For a time-reversal odd order parameter $\chi$, possibly with multiple components, to couple with  $\vec H$ in the free energy $F$, the coupling must involve an odd power of $\vec H$. 
At the same time, if the coupling is linear in $\vec H$, e.g. $F\sim a\cdot \chi_x \cdot H_x$, where $a$ is the coupling constant, the order parameter $\chi$ would directly contribute to the magnetic moment $M\sim\frac{\partial F}{\partial H}$. 
If this is the case, hysteresis in the magnetic moment should be present, inconsistent with the experimental findings. 
To explain the absence of hysteresis in the magnetization data, one needs to consider the higher-order coupling. 
The only symmetry-allowed higher-order coupling that has a distinct symmetry from the linear-order coupling is given by $F\sim b\cdot \chi H_x H_y H_z$, in which case $\chi$ is a scalar $A_g$ under the $D_{2h}$ point-group, but time-reversal odd. 
This order parameter potentially explains the absence of hysteresis in the magnetic moment. 
At the same time, this coupling indicates that a strong angle-dependence in the hysteresis of magnetoresistance, which is not observed in experiments.

% %%%%%%%%%%%%%%%%%%%%%%%%%%%%%%%%%%%%%%%%%%%%%%%
% % APPENDIX
% %%%%%%%%%%%%%%%%%%%%%%%%%%%%%%%%%%%%%%%%%%%%%%%
% \appendix

% \section{Experimental Details}\label{sec:APP experimental}
% \FCB\ crystals were grown using a chemical vapor transport technique from stoichiometric ratios of \FC\ and \FB\ powders (99.9\% purity).
% The starting materials (total mass of 300 mg) were mixed and loaded inside evacuated quartz tubes, which were heated in a three-zone tube furnace to 620~\Cels\ in all 3 zones for 3 days at a rate of 3~\Cels/min. 
% For $x\ge 0.4$, the \FCB\ crystals grew with compositions close to the nominal ones, but for $x< 0.4$, the grown crystals had noticeably higher Cl content than expected. 
% The composition of each sample was determined using energy dispersive x-ray spectroscopy (EDX) in an FEI Scios DualBeam electron microscope equipped with an Oxford Instruments detector.

%The \nocite command causes all entries in a bibliography to be printed out
%whether or not they are actually referenced in the text. This is appropriate
%for the sample file to show the different styles of references, but authors
%most likely will not want to use it.
%\nocite{*}

% \bibliography{References}% Produces the bibliography via BibTeX.

%\end{document}
%
% ****** End of file apssamp.tex ******

\end{addendum}

\end{document}